\documentstyle[graphicx,graphics,hhline,natbib2,natbibmnfix,amssymb,
	amsmath,amsopn,lscape,epsfig]{mn}

\bibpunct[,]{(}{)}{;}{a}{}{}
 
\newif\ifAMStwofonts

\ifoldfss
  \newcommand{\rmn}[1] {{\rm #1}}

  \ifCUPmtlplainloaded \else
    \NewTextAlphabet{textbfit} {cmbxti10} {}
    \NewTextAlphabet{textbfss} {cmssbx10} {}
    \NewMathAlphabet{mathbfit} {cmbxti10} {} 
    \NewMathAlphabet{mathbfss} {cmssbx10} {} 
  \fi
  \ifAMStwofonts
    \ifCUPmtlplainloaded \else
      \NewSymbolFont{upmath} {eurm10}
      \NewSymbolFont{AMSa} {eurm10}
      \NewMathSymbol{\upi}     {0}{upmath}{19}
      \NewMathSymbol{\umu}     {0}{upmath}{16}
      \NewMathSymbol{\upartial}{0}{upmath}{40}
      \NewMathSymbol{\leqslant}{3}{AMSa}{36}
      \NewMathSymbol{\geqslant}{3}{AMSa}{3E}

    \fi
  \fi
\fi 

\ifnfssone
  \newmathalphabet{\mathit}
  \addtoversion{normal}{\mathit}{cmr}{m}{it}
  \addtoversion{bold}{\mathit}{cmr}{bx}{it}
  \newcommand{\rmn}[1] {\mathrm{#1}}

  \newmathalphabet{\mathbfit} 
  \addtoversion{normal}{\mathbfit}{cmr}{bx}{it}
  \addtoversion{bold}{\mathbfit}{cmr}{bx}{it}
  \newmathalphabet{\mathbfss} 
  \addtoversion{normal}{\mathbfss}{cmss}{bx}{n}
  \addtoversion{bold}{\mathbfss}{cmss}{bx}{n}
  \ifAMStwofonts
    \ifCUPmtlplainloaded \else
      \UseAMStwoboldmath
      \makeatletter
      \new@mathgroup\upmath@group
      \define@mathgroup\mv@normal\upmath@group{eur}{m}{n}
      \define@mathgroup\mv@bold\upmath@group{eur}{b}{n}
      \edef\UPM{\hexnumber\upmath@group}
      \new@mathgroup\amsa@group
      \define@mathgroup\mv@normal\amsa@group{msa}{m}{n}
      \define@mathgroup\mv@bold\amsa@group{msa}{m}{n}
      \edef\AMSa{\hexnumber\amsa@group}
      \makeatother
      \mathchardef\upi="0\UPM19
      \mathchardef\umu="0\UPM16
      \mathchardef\upartial="0\UPM40
      \mathchardef\leqslant="3\AMSa36
      \mathchardef\geqslant="3\AMSa3E
    \fi
  \fi
\fi 

\ifnfsstwo
  \newcommand{\rmn}[1] {\mathrm{#1}}

  \DeclareMathAlphabet{\mathbfit}{OT1}{cmr}{bx}{it}
  \SetMathAlphabet\mathbfit{bold}{OT1}{cmr}{bx}{it}
  \DeclareMathAlphabet{\mathbfss}{OT1}{cmss}{bx}{n}
  \SetMathAlphabet\mathbfss{bold}{OT1}{cmss}{bx}{n}
  \ifAMStwofonts
    \ifCUPmtlplainloaded \else
      \DeclareSymbolFont{UPM}{U}{eur}{m}{n}
      \SetSymbolFont{UPM}{bold}{U}{eur}{b}{n}
      \DeclareSymbolFont{AMSa}{U}{msa}{m}{n}
      \DeclareMathSymbol{\upi}{0}{UPM}{"19}
      \DeclareMathSymbol{\umu}{0}{UPM}{"16}
      \DeclareMathSymbol{\upartial}{0}{UPM}{"40}
      \DeclareMathSymbol{\leqslant}{3}{AMSa}{"36}
      \DeclareMathSymbol{\geqslant}{3}{AMSa}{"3E}
    \fi
  \fi
\fi 

\ifCUPmtlplainloaded \else
  \ifAMStwofonts \else 
    \def\upi{\pi}
    \def\umu{\mu}
    \def\upartial{\partial}
  \fi
\fi
	
\def\lcdm{$\Lambda$CDM }

\newcommand{\hkpc}{\mbox{$h^{-1}$ kpc} }

\newcommand{\kms}{\mbox{km s$^{-1}$} }

\newcommand{\kmsDot}{\mbox{km s$^{-1}$.} }

\newcommand{\kmsKet}{\mbox{km s$^{-1}$)} }

\newcommand{\kmsKD}{\mbox{km s$^{-1}$).} }
\newcommand{\Mpc}{\mbox{Mpc} }
\newcommand{\MpcDot}{\mbox{Mpc.} }
\newcommand{\MpcCom}{\mbox{Mpc,} }
\newcommand{\kpc}{\mbox{kpc} }

\newcommand{\msun}{\mbox{$M_{\odot}$} }

\newcommand{\Mrvir}{\mbox{$m_{\rmn{vir}}$} }
\newcommand{\Rrvir}{\mbox{$r_{\rmn{vir}}$} }

\newcommand{\gadget} {{\sc gadget }\ }

\def\la{\mathrel{\hbox{\rlap{\hbox{\lower4pt\hbox{$\sim$}}}\hbox{$<$}}}}
\def\ga{\mathrel{\hbox{\rlap{\hbox{\lower4pt\hbox{$\sim$}}}\hbox{$>$}}}}

\def\gsim{\ga}
\def\lsim{\la}

\newcommand{\bc}{\begin{center}}
\newcommand{\ec}{\end{center}}

\newcommand{\be}{\begin{equation}}
\newcommand{\ee}{\end{equation}}

\newcommand{\tx}[1] {\rmn{#1}}

\newcommand{\zmerg} {$z_{\tx{merg}}$}		
\newcommand{\CGOne} {$CG_{1}$}
\newcommand{\CGTwo} {$CG_{2}$}
\newcommand{\CDMOne} {$CDM_{1}$}
\newcommand{\CDMTwo} {$CDM_{2}$}

\title{On the formation of cold fronts in massive mergers}
\author[H. Mathis, G. Lavaux, J. M. Diego, J. Silk]{H. Mathis$^{1}$\thanks{hxm@astro.ox.ac.uk}, G. Lavaux$^{1,2}$, 
J. M. Diego$^{1}$, J. Silk$^{1}$
\\
$^{1}$University of Oxford, Astrophysics, Denys Wilkinson Building, Keble Road, Oxford OX1 3RH, UK \\
$^{2}$Ecole Normale Sup\'erieure de Cachan, Avenue du Pr\'esident Wilson, 94230 Cachan, France}

\begin{document}
\maketitle
\label{firstpage}



\begin{abstract}

Using adiabatic hydrodynamical simulations, we follow the evolution 
of two symmetric cold fronts developing in the remnant of a violent $z=0.3$ massive cluster
merger.  The structure and location of the simulated cold fronts are very similar to those recently
found in X-ray cluster observations, supporting the merger hypothesis for the origin of
at least some of the cold fronts.  The cold fronts are preceded by an
almost spherical bow shock which originates at the core and disappears
after 1.6 Gyr. The cold fronts last longer and survive until $z\sim0$. 
We trace back the gas mass constituting the fronts and find it initially associated with the two
dense cores of the merging clusters. Conversely, we follow how the
energy of the gas of the initial merging cores evolves until $z=0$ from
 the merging and show that a fraction of this gas can escape from the local
potential well of the sub-clumps. This release occurs as 
the sub-clumps reach their apocentre in an eccentric orbit and 
is due to decoupling between the dark matter and part of the gas in the  
sub-clump because of, first, heating of the gas at first close core passage 
and of, second, the effect of the global cluster pressure which peaks as the 
centrifugal acceleration of the sub-clump is maximal. The fraction of the 
 gas of the sub-clump liberated in the outbound direction 
then cools as it expands adiabatically 
and constitutes the cold fronts. 

\end{abstract}




\section[]{Introduction}
\label{sec:Intro}

     High resolution X-ray temperature images obtained by
\emph{Chandra} of the intracluster medium (ICM) show that large, cold
regions in pressure equilibrium with their surrounding medium are a 
very common phenomenon in massive clusters 
(\citealt{Mark00}, but also \citealt{Vik01,Maz01,Sun02,Mark01a}).  
Their high gas density largely compensates for the factor 
of 1.3 to 4 drop in temperature so that their X-ray surface brightness 
is still typically higher than the rest of the ICM.  The whole cold region is commonly referred
to as a ``cold front'', even if the term would strictly only
describe a possible upstream contact interface of the cold gas with the hot
ICM.

Employing the term in its loose sense, these cold fronts are extended
structures with sizes of one hundred to a few hundred kpc, observed up to a
 significant fraction of the virial radius; X-ray maps show that cold fronts are 
preceded by bow shocks \citep{Vik02,Mark02a}.   
They differ from very dense clumps of gas orbiting the ICM which extend to a few tens of kpc at
most. These clumps are much more difficult to observe except in the central regions
of clusters \citep{Fu02}, and they are expected to be the direct
 remnants of the cores of massive haloes that have merged with the
cluster.  However, cold sub-clumps would share two similar
characteristics with cold fronts: a high X-ray surface brightness
and pressure equilibrium with the ICM.

Numerical simulations have already addressed the formation of
cold fronts: in a pioneering simulations of merging  
clusters of galaxies using Eulerian hydrodynamics, \citet{Roett97} 
 studied cluster-subcluster idealised mergers, and showed that substructure, 
shocks and adiabatic cooling could result in a complex temperature map.

More recently, \citet{Rick01} have addressed the evolution of the X-ray 
luminosity and temperature of the ICM in a series of off-axis mergers of idealised subclusters. 
They show that the  overall luminosity and temperature  increase due to  
merging shocks and regions of gas compression which are observationally unresolved 
can strongly bias the cluster mass if it is derived assuming a non-merging cluster.   
 
\citet{Naga} and \citet{Bia02} have shown that 
cold fronts like those found by \citet{Mark00}
 can form as a result of massive mergers in a cosmological setting. 
While both clearly stated that their simulated 
cold fronts occur when ``the subcluster gas strays from its
local potential minimum and expands adiabatically'' they did 
not explain the process in detail.  Nevertheless, they pointed out that further
observations and simulations need to be carried out to assess how
frequently cold fronts may occur in massive mergers. In fact, massive
mergers need not be the only process responsible for cold fronts, but
they may be sufficient even at low redshift to explain their observed
abundance, all the more if they can remain stable within the ICM for a
sufficiently long time. In a recent paper, \citet{He03} have considered 
ram pressure stripping as a possible mechanism  
for forming some of the cold fronts. They do not include  
dark matter in their simulations, but study the partial unbinding of a dense clump of gas 
undergoing stripping from an external, uniform wind
 switched on at the beginning of the simulation, representing the surrounding ICM. 
This process may be complementary but is not identical to the one we will focus on. 

In the present paper, we study the occurrence of cold fronts as a possible 
result of a major merger, focusing on the interplay 
between the dark matter and the gas of the subcluster cores 
orbiting in the newly formed ICM.  Like 
\citet{Naga} and \citet{Bia02}, we follow the late formation of a  
massive cluster  in the \lcdm cosmology using hydrodynamical adiabatic
simulations. We employ an entropy-conserving DM+SPH scheme
which is suitable to follow the Lagrangian evolution of the flow and the
history of the gas of the cold front. We set the initial conditions to merge
two equal mass haloes into a massive, but plausible, $1.4\times10^{15} \msun$ final
object to more easily bring out the interesting features.

Our massive merger takes place at $z=0.3$. It produces by $z=0.2$ 
two cold fronts which develop in a 500~kpc thick plane containing the orbits of the
two merging cores and expand in the wake of an almost spherically
symmetric ``bow shock''.  This bow shock results from the transonic motion
of the merging subcluster cores in the ICM. The medium may have been previously
partly heated by a merger shock due to the compression of the gas between the
subcluster cores in their convergent motion before their first encounter. 
The dense gas cores of the two merging halos survive until close 
to the present time, when they eventually merge.  As 
the cores orbit in the merger remnant, the gas 
within them is first strongly heated up adiabatically at pericentre by the variation of
the gravitational potential due to close encounter, it then cools down partly 
though not symmetrically between core passage and apocentre as the potential increases again, 
before it is finally heated up by the final inspiralling. 
 
While a fraction of gas particles of the sub-clumps is tidally
stripped at the first closest encounter, both cold fronts form later, 
at the apocentre of the sub-clumps orbits. 
We show that the fronts consist of some of those gas particles
initially within the dense cores which can escape 
out of the local gravitational potential of
the sub-clumps. In our simulation, the cold front phenomenon takes place as the gas of the dense cores
is in slight phase advance relative to the dark matter along the outbound part of
the orbit, and phase delay relative to the dark matter along the inbound
part. At apocentre in an eccentric orbit, when motions are slower but centrifugal acceleration is high, 
the overall pressure inside the cluster acts most strongly, in a direction opposite to the gravitational pull. 
This introduces differences between the dynamical evolution of the gas of the sub-clump and 
that of the dark matter. They are sufficient to result in liberation of gas which 
has been compressed and heated at core passage but has not cooled down symmetrically since then. 
As the cold dense gas leaves the local gravitational potential with high pressure, 
it expands adiabatically and cools down further. Gas which is expelled in the direction of motion 
constitutes the cold front. This gas is not later recaptured by the local gravitational potential 
of the dark matter subclump.

We confirm that the relatively small fraction of the gas mass leaving the local
gravitational potential by this process can make the presence of cold
fronts in X-ray images very dependent on the mass of the
sub-clumps, if cold fronts are mainly produced by the process we describe here. 
We find that the lifetime of our simulated cold fronts 
 exceeds the time required for the bow shock to travel to the virial radius.  

This paper is organised as follows: in Section~\ref{sec:Simus} we
present the simulations. In Section~\ref{sec:Evol} we discuss 
qualitatively the evolution of post-merger physical quantities 
of the ICM as they are directly available from the simulations. 
Sections~\ref{sec:Shocks} and~\ref{sec:ColdFront} 
tackle quantitative aspects of the development of the two main features
apparent on the maps: first, merger shocks and bow shocks, and
second, cold fronts. Section~\ref{sec:PhysPic} focuses on one of the two
cold fronts, proposes a physical explanation for its origin, and briefly 
compares to other work. We summarise and 
conclude in Section~\ref{sec:CCL}. 



\section[]{Obtaining a massive cluster merger}
\label{sec:Simus}

\subsection{Initial conditions}
\label{sec:Simus:IC}

We assume a concordance \lcdm cosmology with parameters $\Omega_0 =
0.3$, $\Omega_b\:h^{2} = 0.019$, $\Lambda = 0.7$, $h = 0.7$.  To study
a major merger, as an alternative to resimulating at higher resolution a
cluster selected from a low resolution simulation, 
we directly constrain the initial gaussian density field so
that it has the usual CDM power spectrum but encapsulates information
on where and how the $z=0$ target cluster should form. For this
purpose we use the \citet{Wey96} implementation of the Hoffman-Ribak
algorithm.   We want an equal-mass massive merger to take
place at $z\simeq 0.5$; this is obtained if we constrain 
the initial density field so that
it has two gaussian peaks of standard deviation $r=5~$\MpcCom with
the same amplitude $A=5.3\times\sigma_{\rmn{r}}$ (where $\sigma_{\rmn{r}}$ is the
root mean square value of the unconstrained initial density field when
smoothed with a gaussian kernel of standard deviation $\rmn{r}$) and
separation $s=27~$\MpcDot Note that constraining the initial
velocities of these peaks adds but little control over the final
merger at the expense of additional tuning, so we have only used the
densities.  The above setup results in a very massive, $2\times10^{15}\msun$ 
cluster at $z=0$ which follows an NFW density profile,
and which formed in an equal-mass merger of two Virgo-sized clusters at
$z_{\tx{merg}}=0.3$. Note that $z_{\tx{merg}}$ is the epoch when the
friends-of-friends groupfinder with linking length 0.2 times the mean
DM interparticle separation merges the two clusters.

\subsection{Simulation details}
\label{sec:Simus:details}

We employ $2 \times 128^3$ {\sc DM} and {\sc SPH} particles in a 100
\Mpc box and start the simulation at $z_{\tx{start}}=100$. 
The {\sc DM} and {\sc SPH} particle masses and
softening lengths read $M_{\tx{DM}}=7.3\times10^{8} \msun$ and
$M_{\tx{SPH}}=1.3\times10^{8} \msun$, and 
$r_{\tx{soft, DM}}=r_{\tx{soft, SPH}}=40\, \kpc$ respectively.  
The softening lengths are kept
fixed throughout in comoving coordinates.  The equations of motion are
integrated with the public version of \gadget \citep{SprGadget2001}
\footnote{\tt http://www.mpa-garching.mpg.de/$\sim$volker/gadget} which we modified to
include the ``standard'' entropy conservation scheme suggested by
\citet{SprEntropy02}. The gas is single-phase, monoatomic with
adiabatic index $\gamma=5/3$ and obeys a perfect gas equation of
state. To estimate the sound speed of the ICM, we assume a 
mean particle mass $\bar{\mu}\,m_{\rmn{p}}$ with $\bar{\mu}=0.59$ 
and $m_{\rmn{p}}$ is the proton mass. We do not include cooling, 
heating, thermal conduction, magnetic fields, 
nor mechanical or thermal AGN/supernova feedback as
our focus is the capability of a massive merger to induce large, long-lasting 
cold fronts even in an adiabatic evolution. We focus on 11 outputs
equally spaced between $z_{\tx{merg}}$ and $z=0.1$, and 3 outputs at
$z=0.06$, $z=0.03$ and $z=0$.  The final 
virial mass and radius of the cluster, where the enclosed DM
density drops to 200 times the critical density, are 
$\Mrvir=1.35\times10^{15} \msun$ and $\Rrvir=3.4$~Mpc.

\begin{figure*}
  \includegraphics[angle=270,width=\columnwidth]{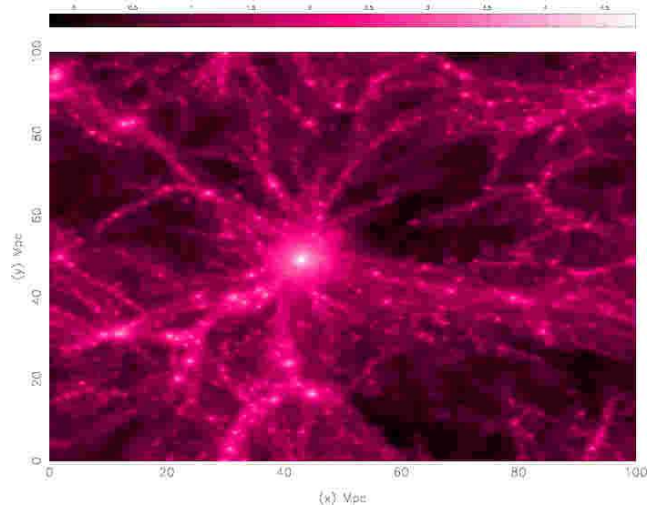}
	  \caption{\label{fig:Fig1} 
Projected $z=0$ gas density of the simulation. The region shown is 100 Mpc wide and we 
have projected a 10 Mpc thick slice normal to the $z$ direction and centred on the cluster. 
The densities are colour-coded
logarithmically. The cluster hosting the cold fronts, at the centre of the image, 
appears well relaxed, but note that the spatial and dynamic range 
chosen here make it difficult to fully render the substructure of the cluster.  
A better, high resolution version of this figure is available at: {\tt http://www-astro.physics.ox.ac.uk/$\sim$hxm/ColdFronts/}.}
\end{figure*}

Figure~\ref{fig:Fig1} gives the projected DM density of the whole box, 
with the cluster at the centre. The size of the region shown is 100~Mpc, and the whole 
box has been projected. The two massive subclusters 
merging at $z=0.3$ infall along the diagonal filament spanning 
from the lower left to the upper right. Once it has formed, 
the cluster still accretes smaller clumps of DM, mostly from the filament.  

\subsection{Tracking the cores of the merging subclusters}
\label{sec:Simus:TrackingCore}

\begin{figure}
  \includegraphics[width=\columnwidth]{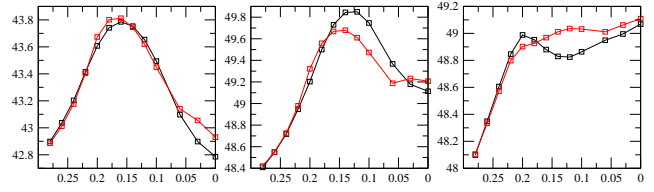}
  \caption{\label{fig:Fig2}The black and red lines show the
    spatial evolution after $z_{\tx{merg}}=0.3$ of the centre of mass of
     the gas and dark matter (respectively) of the core of the subcluster
    which was accreted from the lower left of the simulation (\CGOne, \CDMOne). 
    Left, middle and right panels corresponds to the x-, y- and z-positions.  Note how the centre of
    mass of the gas is offset to the outer parts of the cluster 
    with respect to the centre of mass of the DM in y coordinates at
    $z=0.12$ by $\sim200$ kpc: the dark matter has already fallen back
    towards the centre while the gas clump is still stationary: the upper
    right cold front appears at this point. Note also that the $z\sim0.2$ delay of the 
    centre of mass of the gas with respect to that of the 
    dark matter may be a consequence of the pressure gradients as 
core passage takes place then.}
\end{figure}

\begin{figure}
  \includegraphics[width=\columnwidth]{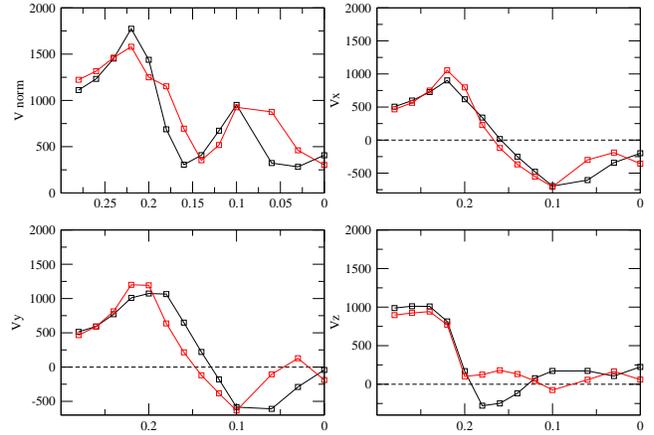}
  \caption{\label{fig:Fig3} The black and red lines show the 
 evolution after $z_{\tx{merg}}=0.3$ of the velocity of the centre of 
mass of the gas and dark matter respectively of the core of the subcluster 
accreted from the lower left of the simulation. The top left panel shows 
the amplitude of the 3D velocity, while the upper right, lower left and lower 
right panels respectively give $v_x$,  $v_y$ and $v_z$. Velocity is 
transonic at $z\sim 0.21$. Negative accelerations along the 
x and y directions peak at $z\sim0.17$.}	
\end{figure}

\begin{figure}
  \includegraphics[width=\columnwidth]{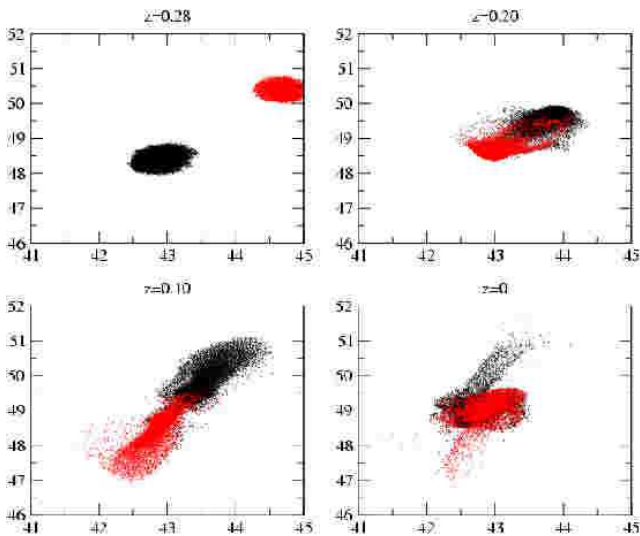}
  \caption{\label{fig:Fig4} 
Positions of the gas particles of the clumps \CGOne$\,$ (black) and \CGTwo$\,$ (red)
at four different redshifts in the x-y plane. The particles 
have been selected using a friends-of-friends groupfinder 
and their evolution was tracked down. The window is the 
same for the four plots. 
A better, high resolution version of this figure is available at: {\tt http://www-astro.physics.ox.ac.uk/$\sim$hxm/ColdFronts/}.}
\end{figure}

The  merger occurring at $z=0.3$ involves two subclusters of
fairly similar, Virgo-size masses. Following \citet{Bia02}, 
we select their two cores (both gas and DM) 
before $z_{\tx{merg}}$, using a linking length $0.05$ times
 the mean interparticle separation. We find that the trajectory of the
 cores of the subclusters is roughly confined to a constant $z$ slab with
 thickness 500~kpc, from $z_{\tx{merg}}$ down to $z=0$ where
 they completely mix. Figure~\ref{fig:Fig4} shows the particles of the two 
gas cores projected along the z-direction, over a series of snapshots. 
The cores have a first close encounter
 with small impact parameter at $z=0.2$, reach their apocentre, and
 finally mix up at $z=0$. We label as subcluster 1 and 2 the halos
 approaching at $z_{\tx{merg}}$ from the lower left corner (in black) and
 upper right (red) corners respectively. In the following discussion,
 we will refer to \CGOne$\,$ (resp. \CDMOne) and \CGTwo$\,$ (\CDMTwo) as the gas (DM)
  of the two subcluster cores. The resulting number of
 particles and masses of the two cores are $N_{\tx{DM,1}}=12500$,
 $N_{\tx{SPH,1}}=10200$, $M_{\tx{DM,1}}=1.5\times10^{14} \msun$,
 $M_{\tx{SPH,1}}=1.7\times10^{13} \msun$ and $N_{\tx{DM,2}}=10500$,
 $N_{\tx{SPH,2}}=7200$, $M_{\tx{DM,2}}=0.9\times10^{14} \msun$,
 $M_{\tx{SPH,2}}=0.96\times10^{13} \msun$ respectively. Figs.~\ref{fig:Fig2} 
and~\ref{fig:Fig3} respectively give the position and velocity 
of the centres of mass of the gas and dark matter component 
of the core of the subcluster accreting from the lower left of 
the simulation (\CGOne$\,$ and \CDMOne$\,$ correspond 
to the black and red curves). Note in Fig.~\ref{fig:Fig2} how
 the centres of mass of the two components of the sub-clump 
are confined to a thin slab normal to the z-direction.  
Note also the strong deceleration in $v_{\rmn{x}}$ and  
$v_{\rmn{y}}$  seen around $z=0.17$ in the top 
right and bottom left panels of Fig.~\ref{fig:Fig3}, 
as the clump reaches the apocentre if its orbit. 

\section[]{Post-merger evolution of the ICM}
\label{sec:Evol}

In this Section, we track the post-merger evolution of the ICM using
maps of a $6\times6$ Mpc slice cut in the x-y plane of the simulation and
centred on the most bound DM particle of the cluster at $z=0$. Here and
in all the following the spatial analysis of the plots and the discussion 
are in comoving length. The slice is 500~kpc thick and also centred in the $z$
direction. All quantities are projected along the line of
sight, {\it i.e.} in the z-direction. Due to the small thickness of the slice,
we expect only minor blurring because of projection effects.  While
temperature and pressure fully describe the thermodynamics of the gas,
both observable as respectively, the X-ray emission-weighted
temperature and the Compton $y$ parameter, the specific entropy,
although not directly observable enables one to trace non-adiabatic 
phenomena like shocks and has become widely used in studies of the ICM.

\begin{figure*}
  \includegraphics[width=\textwidth]{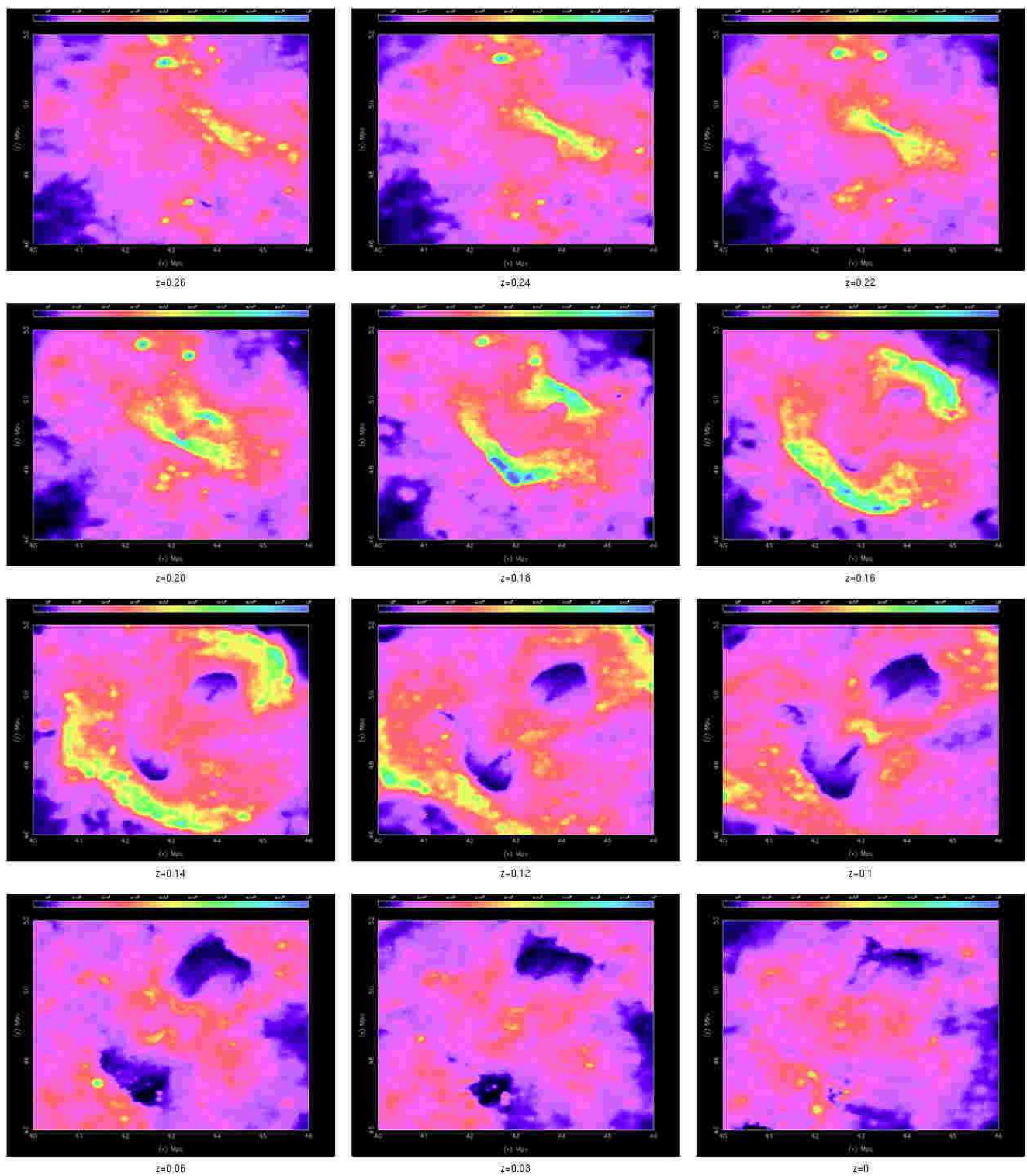}
  \caption{\label{fig:Fig5} Series of 
snapshots, from $z=0.26$ (top left panel) to $z=0$ (bottom right panel) of the X-ray emission-weighted 
temperature integrated over a 500~kpc thick slice. Each image shows the same $6\times6$ Mpc region 
in the x-y plane. The temperatures are colour-coded linearly and the palette 
is chosen to emphasise the luminosity contrast for low temperatures. 
The colour scale ranges from $10^8$ to $10^9$ K and it is the same for all snapshots.
A better, high resolution version of this figure is available at: {\tt http://www-astro.physics.ox.ac.uk/$\sim$hxm/ColdFronts/}.}
\end{figure*}

\begin{figure*}
  \includegraphics[width=\textwidth]{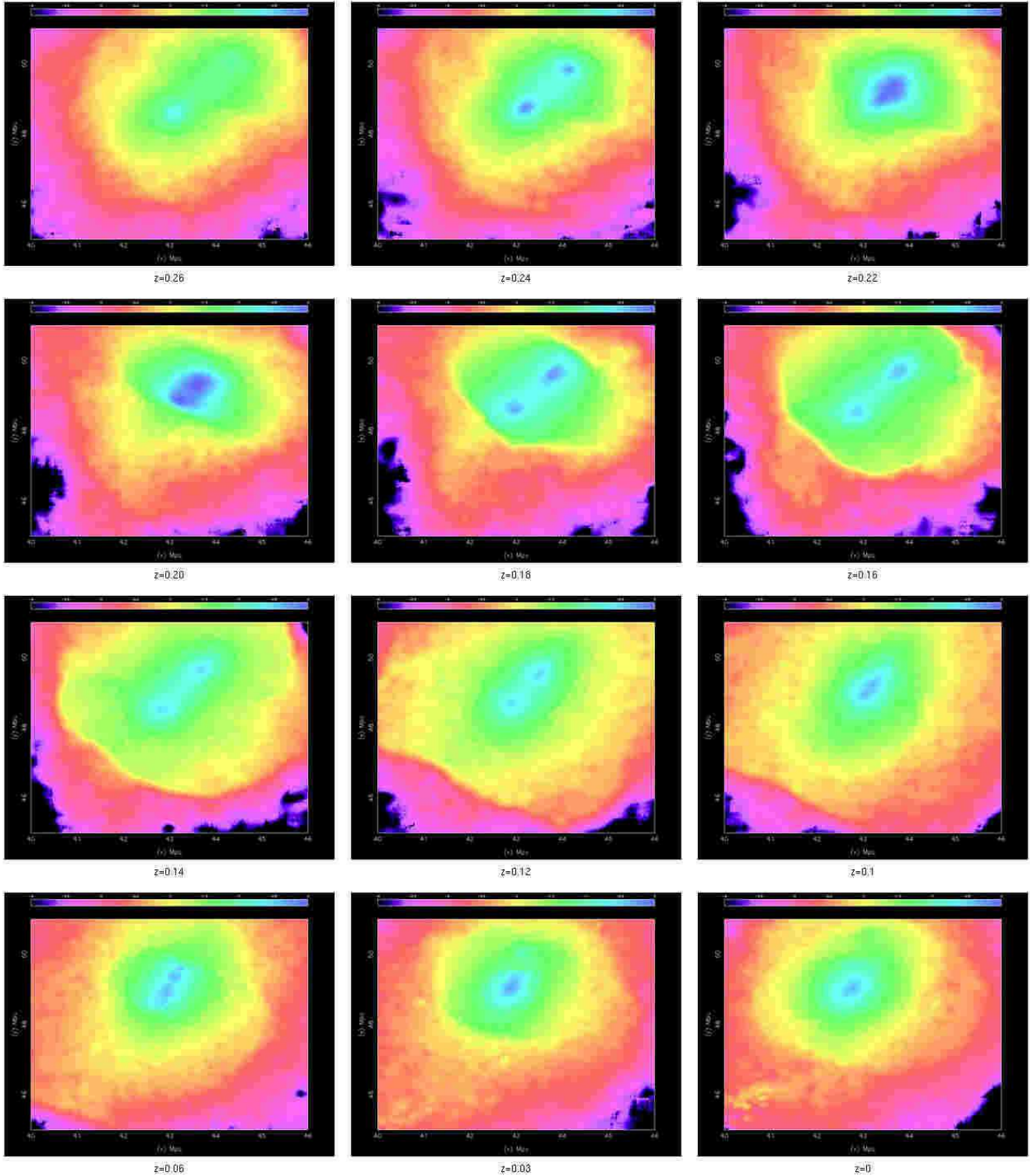}
  \caption{\label{fig:Fig6} Same as 
Fig.~\ref{fig:Fig5}, but for the Compton $y$ parameter 
of the gas integrated over the 500~kpc thick slice. The colour scale is 
logarithmic and the normalisation is arbitrary.
A better, high resolution version of this figure is available at: {\tt http://www-astro.physics.ox.ac.uk/$\sim$hxm/ColdFronts/}.}
\end{figure*}

\begin{figure*}
  \includegraphics[width=\textwidth]{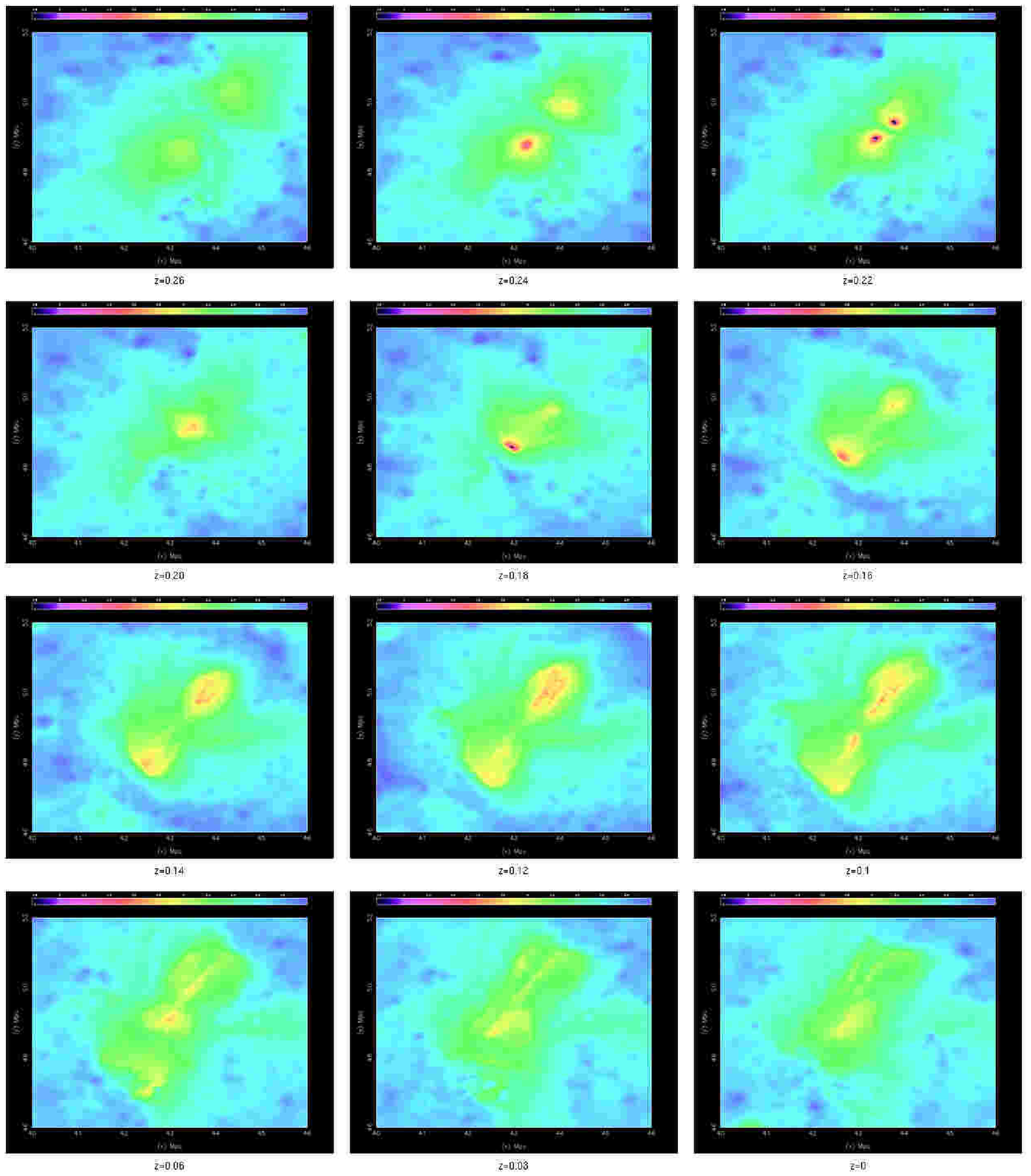} 
  \caption{\label{fig:Fig7} Same as Fig.~\ref{fig:Fig5}, but for the decimal logarithm 
of the adiabat A(s). The colour scale is here a linear function of the specific entropy of the gas.
A better, high resolution version of this figure is available at: {\tt http://www-astro.physics.ox.ac.uk/$\sim$hxm/ColdFronts/}.}
\end{figure*}

Figures~\ref{fig:Fig5},~\ref{fig:Fig6} 
and~\ref{fig:Fig7} show consecutive maps of, respectively, the
bremsstrahlung-emission-weighted temperature of the ICM, the Compton
$y$ parameter and, following common practice, the adiabat $A(s)$ of
the gas defined as $P=A(s)\,\rho^{\gamma}$.  $A(s)$ is a monotonic
function of the specific entropy of the gas, and it is related to the
formal thermodynamic entropy per particle $s$ as: $A(s)=e^{2/3\;
(s-s_{0})}$ where $s_{0}$ depends only on fundamental constants and
on the mixture of particle masses \citep{Voi03}.  We divide the full
sequence in three consecutive phases: first, the initial compression
of the intervening gas, between $z_{\tx{merge}}$ and $z=0.22$, second,
the generation of the cold fronts, from $z=0.2$ to $z=0.12$, and
third, the final relaxation from $z=0.1$ to $z=0$.  We only give a
qualitative description and we postpone the quantitative
analysis to the following two sections.  To conclude this section, we
give X-ray luminosity maps in four typical configurations to show 
resemblance to the observations.

\subsection[]{Compression of the intervening gas}
\label{sec:Evol:CentralComp}

The first four panels of Figs.~\ref{fig:Fig5} to~\ref{fig:Fig7} 
show how the intracluster gas between the 
two subcluster cores is compressed and shock heated as they approach 
the centre from the lower left and upper right corners respectively.  The
compression at the centre of the forming cluster is
obvious in the temperature maps (Fig.~\ref{fig:Fig5})
where a slab of hot gas normal to the axis of merging of the
subcluster builds up and starts to expand. The top panel 
of figure~5 of \citet{Mark00} presents a schematic picture 
of the first stage of a massive merger that they use to model their  \emph{Chandra} 
observation of A2142. (This picture was first derived    
from simulations by \citealt{Roett97} and later confirmed by \citealt{Rick01}.)  
  Following their view, the compression we find would 
eventually lead to the formation of a merger shock. At $z=0.22$, the dumbbell
pattern visible on the temperature maps is a direct consequence of the dense
cool cores compressing the hot gas. Note that
Fig.~\ref{fig:Fig7} also clearly shows at $z=0.22$ the
location of the subcluster cores as they host the lowest entropy gas
of the map. Note also that the absence of subcluster cores in the
$z=0.26$ panels of the Compton $y$ parameter and of the entropy
(Figs.~\ref{fig:Fig6} and~\ref{fig:Fig7})
is simply due to the two cores being slightly out of the slice chosen for the
maps: they reappear by $z=0.24$.

In our simulation, merger shocks would take place in the ICM as the 
subcluster gas surrounding one core and carried out in its motion hits 
the gas of the other subcluster with opposite direction of motion and supersonic 
relative velocities. At $z=0.22$, \CGOne $\,$ and \CGTwo $\,$ have velocities of order 1700 km/s 
 with respect to the rest-frame of the simulations 
(twice this for their relative velocity, see the upper left panel of Fig.~\ref{fig:Fig3}). 
 The speed of sound in the unshocked ICM is of the same order as the velocity of one clump 
for a $10^{8}$K ICM (1530 \kmsKet  and higher for $2\times10^{8}$K 
(2160 \kmsKD Shocks are therefore expected to develop; in the following, 
we will refer to these pre-encounter shocks as ``merger shocks''. 
We will see that they are difficult to follow as they soon correlate 
to other phenomena.  
   
\subsection[]{Generating the cold fronts}
\label{sec:Evol:CFDevel}

The merger shocks expand from the central regions into the ICM as the
subcluster cores go past one another. At $z=0.2$, the fourth panel of
Fig.~\ref{fig:Fig5} shows the first stage of
expansion of the merger shocks, propagating at close to the speed of sound in 
the ICM. However, as foreseen by \citet{Mark00} (see the lower panel
of their fig.~5), while the subcluster cores head towards their
apocentre, they create new shocks as they move in the central
cluster regions where the gas has been previously heated by the merger
shocks. We will refer to this new set of shocks as ``bow
shocks''. The bow shocks propagate out at speeds 
set by the bulk velocities of the sub-clumps, 
when their motion is supersonic, and become pressure waves 
propagating at the speed of sound when the motion of the clumps 
in the surrounding ICM becomes subsonic. In fact, they seem to 
catch up with the merger shocks in the $z=0.18$ panel of 
Fig.~\ref{fig:Fig5}, where they interfere 
and make a complex structure. We will consider 
for simplicity that together they can be described 
by only one shock wave, 
and we will not resolve its inner details. In a pattern
similar to a temperature annulus, the shock/pressure wave expands up to the virial radius
until $z=0.12$, and it dissolves later in the far outskirts of the cluster (IGM, filaments) not 
seen on the maps. The birth of the bow shocks is clearly seen in the $z=0.20$ panel of the
Compton $y$ parameter (Fig.~\ref{fig:Fig6}): the steep
pressure rise due to the bow shock appears at the lower left edge of 
the central light blue zone.  The pressure gradient is visible at 
$z=0.18$ and even more at $z=0.16$ where it clearly delineates both 
lower left and upper right shocks (or pressure waves). 
Entropy maps (Fig.~\ref{fig:Fig7}) mainly trace the positions of the
subcluster cores and show how they lose a fraction of their
low entropy gas to the central region of the cluster after their first
crossing, even before reaching the apocentre. The entropy rise due to
shocks is less visible however, because we expect a small increase due
to shocks in the units we express the entropy: a shock with upstream Mach number 
$M=1.4$ (2.2) result in a 1 (10) percent change in $\log{A(s)}$. We will tackle 
more quantitative aspects of the shocks in the next section.

Fig.~\ref{fig:Fig5} shows little temperature
variation between the up- and downstream flows (excluding the broad shock region), 
at $z=0.18$ and at $z=0.16$.  This is due to the adiabatic outward 
expansion of the gas in the downstream region. 
Two symmetric cold fronts develop in the wake of the bow shocks: while
the bottom left front already becomes visible on the temperature map
(Fig.~\ref{fig:Fig5}) at $z=0.18$, both reach
their final position by $z=0.12$, where they strongly resemble the
bow-shaped features of high surface brightness 
observed in the high resolution X-ray images (see,
e.g., the top left panel of figure~4 in \citealt{Mark02b}). Continuity
of the Compton $y$ parameter over the cold fronts is already demonstrated in
Fig.~\ref{fig:Fig6}. The entropy profiles 
(Fig.~\ref{fig:Fig7}) indicate that cold fronts are
primarily made of low entropy gas, and we will probe the origin of this gas
in Sections~\ref{sec:ColdFront} and~\ref{sec:PhysPic}. Note the
persistence of a filament of low-entropy gas joining the two sub-clump
cores from $z=0.16$ down to $z=0.10$. This filament traces continuous
gas accretion on the central region of the gas particles of \CGOne$\,$, \CGTwo $\,$ 
which have been unbound at  first close encounter of the cores and 
of gas which is gradually stripped off the cores as they orbit into the ICM.
 
\subsection[]{Final relaxation}
\label{sec:Evol:FinalRela}

The pressure waves keep on extending out of the frame shown on the maps. 
Their integrity is already broken at $z=0.12$ but some features persist  
down to the present day. To the opposite, the cold fronts remain 
fairly stationary and stable on the temperature maps until $z=0.03$ 
and only weaken by $z=0$. The cold fronts survive the shock preceding them, not only 
because they stall in their outward propagation, but also because 
they do not start to ``wash out'' in temperature maps as the shocks do in our 
simulation on the $z=0.12$ and $z=0.10$ maps, even they it can still be 
tracked later on to larger radii. Therefore, if major mergers are 
the main cause of cold fronts, their smaller duty 
cycle will make the bow shocks they are associated with 
intrinsically more difficult to detect in X-ray temperature maps. 

In the meantime, the sub-clump cores merge 
as they reach their second close encounter: while 
the pressure maps in Fig.~\ref{fig:Fig6} track the final inspiralling and 
merging, the entropy maps of Fig.~\ref{fig:Fig7} 
show a more complicated behaviour. In addition to the final merging, 
they show two streams linking the central region to 
the top right cold front, which 
develop between $z=0.06$ and $z=0.03$.

We stress here that the temperature and Compton $y$ parameter maps of 
Figs.~\ref{fig:Fig5} and~\ref{fig:Fig6} have been obtained on a 500~kpc thick 
slice containing the orbits of the two subcluster cores. This has allowed us 
to achieve high spatial resolution and to precisely visualise interesting features. 
In reality, limited resolution and projection effects will degrade the maps. 
 Recent observations have improved by large factors in resolution 
(\emph{Chandra} reaches 1'', corresponding to a comoving size of 4.6 \hkpc at $z=0.3$), but 
projection effects may still significantly smooth and distort details like those shown 
on our maps. Features which are prominent and asymmetric with respect 
to the cluster centre like cold fronts may be the least affected, 
but thin patterns more symmetric with respect to the cluster centre 
and extending over a larger area such as shocks may 
significantly suffer from projection. Note also that because X-ray emission 
is concentrated at the cluster centre, observations  of temperatures variations 
which are weak and/or with small spatial extent far from the centre present 
a major challenge.

\subsection[]{An X-ray picture of the ICM}
\label{sec:Evol:X-ray}

\begin{figure} 
  \epsfig{file=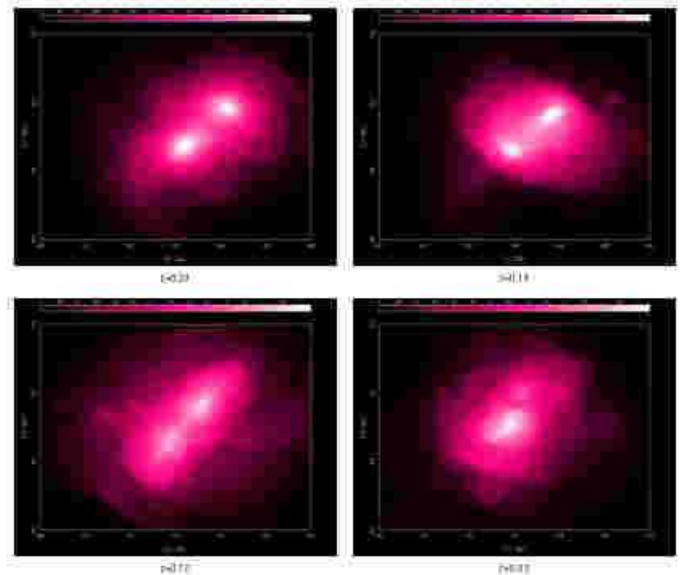,width=9cm}
  \caption{\label{fig:Fig8}
X-ray bolometric surface brightness $S_{X}$ (arbitrary normalisation) at 
four different stages after $z_{\tx{merg}}$. $S_{X}$ is 
colour-coded logarithmically and has been integrated over a 500~kpc-thick 
 slice. The region shown is the same as in Fig.~\ref{fig:Fig5}, and  
the colour scale is the same for the four images. Note that 
the upper right picture is strikingly similar to some of the \emph{Chandra} images in \citet{Mark00,Mark02a}, 
as is the sharp transition between red and black to the lower left of the $z=0.03$ panel, 
due to the cold front. In reality, projection effects may degrade the sharpness of the weak features, 
but the strongest contrasts between white and orange colours will remain unaffected 
(see text for discussion). A better, high resolution version of this figure 
is available at: {\tt http://www-astro.physics.ox.ac.uk/$\sim$hxm/ColdFronts/}.}
\end{figure}

Some of the physical processes mentioned above are directly echoed in
X-ray observations of the ICM. Figure~\ref{fig:Fig8}  gives the
X-ray surface brightness $S_{\rmn{X}}$ at four different times, over the same, 500~kpc-thick 
region that was used for Figs.~\ref{fig:Fig5} to~\ref{fig:Fig7}. The panels 
show a variety of morphologies. 

The top left panel shows two spherical, clearly separated
subcluster cores at $z=0.24$ (in white), before their first close encounter.  At
$z=0.18$ (top right panel), as they orbit towards their apocentres,
the two cores begin to lose a fraction of their gas in their wake, 
seen as diffuse white emission, and the transition from red to black in the
lower left and upper right parts of the image becomes sharper, a
direct consequence of the merging shock propagating outwards. The
resemblance of this panel with some of the \emph{Chandra} pictures
discussed by \citet{Mark02a} or \citet{Mark00} is striking.  At
$z=0.12$ (bottom left panel) the cores contract, acquire an elongated shape, 
and fall back in the central potential well, while the external
pressure wave progresses towards the outer layers of the cluster. The bottom right panel 
($z=0$) shows what seems a fairly relaxed cluster, but with the clear 
luminosity gradient of the cold front of the bottom left region. 

Again, note that projection effects may blur features in the X-ray 
luminosity which are clearly seen when using 
a 500 kpc-thick slice cutting through the cluster centre. In fact, 
 X-ray emission is much more compact and peaked than, 
for instance, the Sunyaev-Zeldovich effect measuring the 
Compton $y$ parameter. This becomes obvious when comparing 
Fig.~\ref{fig:Fig8} to Fig.~\ref{fig:Fig6} at $z=0.18$. In the former, 
the diameter of the inner ``structure'', taken at a factor 
100 down from the peak of the emission, does not exceed 
$\sim$2$\,$\Mpc while the high-pressure zone, 
at the same dynamic range from the peak value, reaches 
more than 4$\,$\Mpc on the $z=0.14$ slice of Fig.~\ref{fig:Fig6}.
As a result, the smoothing due to projection will be much reduced 
in X-ray luminosity maps in comparison to pressure maps. 

\section[]{Resolving shocks in the ICM}
\label{sec:Shocks}

We have only considered one global shock in the previous section, 
although complex multiple shocks in the ICM are expected to develop during mergers. 
In fact, the substantial density of the core of our merging subclusters, together with their high 
infalling velocity results in a variety of supersonic features. 
According to common lore,  SPH simulations only provide limited shock front resolution.  
Here, however, the high resolution of the simulation enables us
to partly circumvent this restriction and to bring out two aspects of shock 
propagation that we find particularly striking. 
We first focus on the $z=0.20$ and $z=0.18$ snapshots, when 
the bow shock or its pressure wave catch up with the merger shock. 
Then, we briefly discuss the possibility of secondary shocks.

\subsection[]{Merger and bow shocks}
\label{sec:Shocks:MegerBow}

First, we discuss the primary quantities temperature 
and pressure profiles across the shock and cold 
front selected at $z=0.14$. Then, we consider 
the velocity field and entropy evolution as secondary probes.

\subsubsection[]{Temperature and pressure profiles}
\label{sec:Shocks:MegerBow:TempPress}

The broad geometry and sequence 
of temperature variations remain simple and seem to confirm the 
picture sketched by \citet{Mark00}: merger shocks first form in a slab 
separating the two subcluster cores and normal to the direction of merging  
as the regions of the gas surrounding the subclusters situated in front 
of each of them  penetrate one another at supersonic velocities. These merger 
shocks propagate outwards as pressure waves in the ICM 
once the motion of the subcluster gas in which they originate 
has slowed down or when the merger shocks have swept
up all the subcluster gas and are situated at the interface between
the shock heated subcluster gas and the relatively more stationary 
ICM with lower opposite velocity.  Bow shocks, on the other hand, are due to 
the supersonic (transonic here) motion of the dense gas 
cores \CGOne$\,$and \CGTwo $\,$of the subclusters inside the newly formed 
ICM. The ICM includes contribution from the subcluster gas 
surrounding the dense cores which has been stripped off by 
the merger shocks. As bow shocks 
can propagate out in the ICM faster than the merger shocks 
or their associated pressure wave if the outbound clump velocity 
remains sufficient, they can eventually merge with the latter. 
The complexity of the symmetric high-temperature regions in Fig.~\ref{fig:Fig5} 
(see $z=0.18$) suggests that this is the case in our simulation. 

\begin{figure}
  \includegraphics[width=\columnwidth]{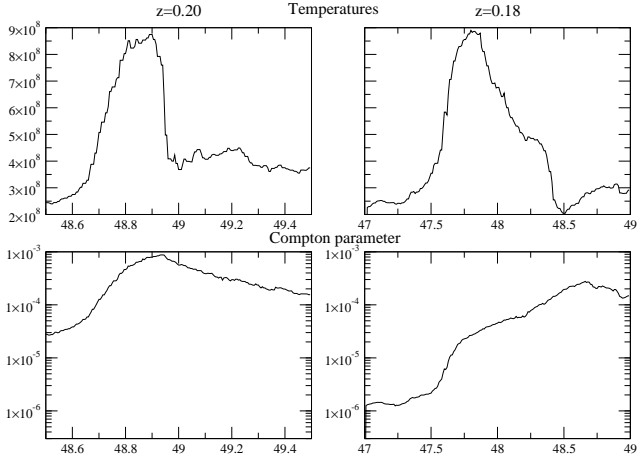}
  \caption{\label{fig:Fig9} 
Temperature and Compton $y$ parameter projected over a 500~kpc 
thick slice in $z$, computed along a test line cutting 
through the lower left bow shock in the x-y plane, 
at $z=0.2$ and $z=0.18$ (see text for details on the choice of the test line). 
The abscissa shows the y-coordinate. The upstream region 
of the shock is to the left of the temperature and pressure rise. 
While the cold subcluster clump is not visible on the $z=0.2$ 
temperature cut, its extension is evident at $y=48.5$ on the $z=0.18$ temperature cut.}
\end{figure}

The left and right top panels of Fig.~\ref{fig:Fig9}
show the temperature profiles across the lower left shocks at $z=0.20$
and $z=0.18$ respectively, while the left and right bottom panels give
the Compton $y$ parameter profiles at the same redshifts.  The
profiles show the projected quantities along a test segment reaching
from the unaffected ICM (at the exterior of the outermost shock front)
in the bottom left region of the above Figures~\ref{fig:Fig5}
to~\ref{fig:Fig7} to deep in the core region of the newly
formed cluster. This line probes the highest temperature zones (in
blue) of the lower left shocks at $z=0.2$ and $z=0.18$. 
Note that because these temperature extrema are drifting in
the x-y plane, we have taken the test line parallel to the y-axis with
the $z=0.20$ lower and upper end points at (43.1, 48.5) and (43.1,
49.5) Mpc and the $z=0.18$ lower and upper end points at (43.1, 47)
and (43.1, 49) Mpc. The abscissa of Fig.~\ref{fig:Fig9} 
is labelled in the global y-coordinate of the simulation (in Mpc). 
This test line setup allows us to capture all the above features 
on a straight line. We note that 
since the test line is not strictly parallel to the direction of 
propagation of the shock, this offset will dilate the scale. A stricter  
result would be obtained using a moving test line so that it remains 
aligned with the direction of propagation of the shock, 
but we would not gain much more information.

At $z=0.20$, the sub-clumps have gone slightly past their first
closest encounter (see the top right panel of Fig.~\ref{fig:Fig4}). 
The bulk of the \CGTwo$\,$ particles reach $y=48.5$ at $x=43.1$, 
but the temperature and pressure rise characterising 
the shock are clearly seen at $y\sim48.7$. Between 
$y=48.65$ and $y=48.8$, the temperature
rises by a factor of $T_{2}/T_{1}\sim2.7$ while the pressure rises by a
factor $P_{2}/P_{1}\sim10$, where indices 1 and 2 respectively label the
upstream and downstream regions. Assuming a flow normal to the shock, 
the Rankine-Hugoniot relations give 
a high upstream Mach number $M_{1}\sim2.6$ to 2.8. 
We will show in paragraph~\ref{sec:Shocks:MegerBow:VelEntropy:Vel} 
below that this value is too high and that 
the upstream velocity of the flow with respect 
to the shock is of order 2000 $\,$\kms only: the 
flow is transonic/slightly supersonic. (If the flow is significantly 
oblique to the high temperature front, it may be subsonic, 
without shock).  Because of this uncertainty, we 
use the entropy evolution in 
paragraph~\ref{sec:Shocks:MegerBow:VelEntropy:Entropy} 
 to verify that a shock actually develops. 

As we have already stressed in the previous section,  
it is difficult to estimate how the 
merger shock, due to the initial compression of the ICM gas intervening between 
the subcluster cores during their first approach, or its remnant 
adds to the effect of the bow shock induced in the direction of motion of the subcluster cores 
after their first close encounter as the steep rise singled out on the 
temperature and Compton $y$ profiles at $z=0.20$ is generated. 
In fact, both types of shocks may result in features similar  to the ones we observe. 
 Clearly separating these two components would require a full 3D analysis 
and even higher spatial and temporal resolution. 

In the downstream region corresponding to the 
central zone of the newly formed cluster, 
the temperature of the gas heated by both shock and compression 
decreases as a result of the adiabatic expansion of the gas in 
the wake of the shocks; this is due to the 
divergent motion of the flow as it follows \CGOne$\,$ and
\CGTwo$\,$. Note that at $z=0.20$ the temperature $T_{2}$ of the flow is
fairly constant over the range $y=49$ to $y=49.5$.  The pressure
$P_{2}$ declines monotonically by a factor of 5 from the peak at
$y=49$ all the way to $y=49.5$.

Later, at $z=0.18$, as the sub-clump cores are already 1.5~Mpc apart
in the x-y plane and on their way to their apocentre, they can be
viewed as two pistons compressing the gas in front of them. Because
their motion is not supersonic anymore in a $2\times10^{8}$K hot ICM, they 
will not sustain the shock. As a result, it transforms into 
a pressure wave propagating upstream. The upper right panel of Fig.~\ref{fig:Fig9} 
indicates that the wave is some 750~kpc ahead of \CGTwo$\,$, at $y=47.6$, 
with the outskirts of the cool core of dense gas clearly visible at $y=48.5$ 
on the temperature profile. This is confirmed by comparison 
to the $z=0.18$ snapshot of Fig~\ref{fig:Fig7}. 

Note that no such temperature drop was visible on the $z=0.20$ map simply because both 
\CGOne$\,$ and \CGTwo$\,$ are then slightly out of the 500~kpc thick slice used to compute
the profiles.  The amplitudes of the jumps in both pressure and density
are similar at $z=0.20$ and $z=0.18$, but the shapes of the profiles differ in the
downstream region, a possible consequence of the readjustment 
of gas and dark matter density profiles in the central region of the cluster.

\subsection[]{Velocity and entropy across the shock}
\label{sec:Shocks:MegerBow:VelEntropy}

\subsubsection[]{Gas velocity}
\label{sec:Shocks:MegerBow:VelEntropy:Vel}

\begin{figure}
  \includegraphics[angle=270,width=\columnwidth]{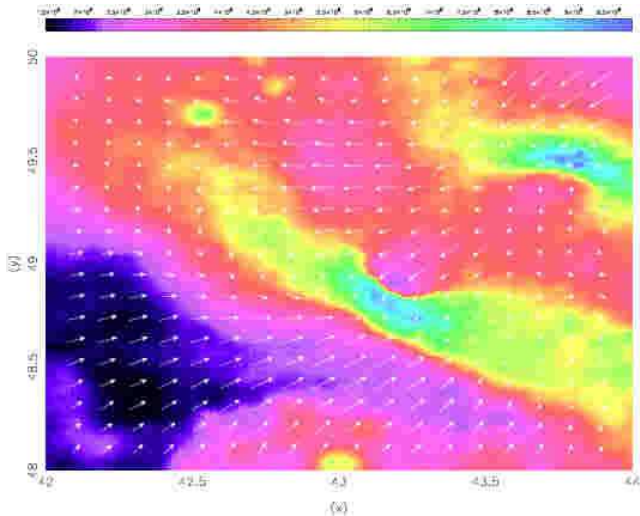}
  \caption{\label{fig:Fig10} The $z=0.2$ projected gas velocity field in a 500~kpc thick, 
2~Mpc wide slice cut in the x-y plane around the lower left cold gas core. The background 
shows the emission-weighted temperature map (colour-coded linearly with a scale particular to this Figure).  
The largest arrows correspond to a velocity of $\sim1000\,\kmsDot$ Note the discontinuity 
 of the velocity field over the shock (for instance, $x=43.25$, $y=48.5$).
A better, high resolution version of this figure is available at: {\tt http://www-astro.physics.ox.ac.uk/$\sim$hxm/ColdFronts/}.}
\end{figure}

\begin{figure}
  \includegraphics[angle=270,width=\columnwidth]{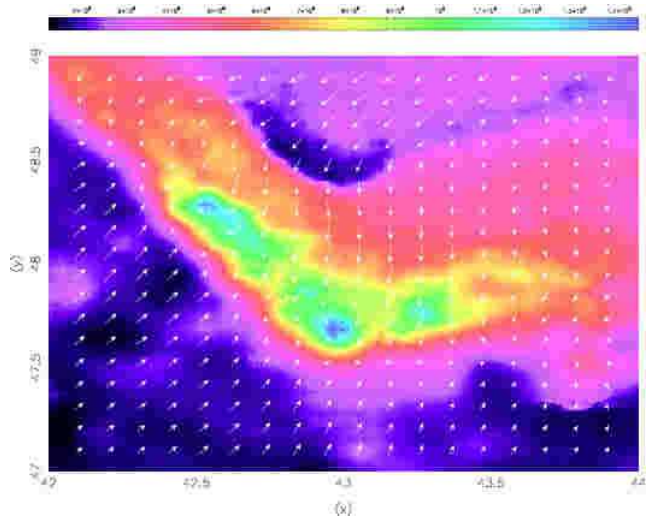}
  \caption{\label{fig:Fig11}Same as Fig.~\ref{fig:Fig10}, 
but at $z=0.18$ and with a different colour scale for the temperature. The region shown has been  translated downwards from 
Fig.~\ref{fig:Fig10} by 1$\,$Mpc, to better follow the shocks. 
A better, high resolution version of this figure is available at: {\tt http://www-astro.physics.ox.ac.uk/$\sim$hxm/ColdFronts/}.}
\end{figure}

Fig.~\ref{fig:Fig10} shows the projected gas 
velocity field at $z=0.2$ of a square slice of width 2 Mpc and 
thickness 500~kpc spanning over the lower left shock region, overlaid 
on the colour-coded emission-weighted temperature map selected from 
the same slice. Note that to best cover the temperatures ranges in 
the selected areas shown here and in Fig.~\ref{fig:Fig11}, 
the temperature colour scales are different in Figs.~\ref{fig:Fig10} 
and~\ref{fig:Fig11}, and that they also differ 
from the scale of Fig.~\ref{fig:Fig5}. 

The velocities have been computed in the rest-frame 
of the simulation, and the longest arrows correspond to 1000 km/s. 
While the ICM outside of the bow shock (to the lower left) is 
accreting towards the cluster centre, the amplitude of the velocity of 
the gas strongly decreases as it hits the strong positive temperature gradient, 
for example over the thin yellow line passing over ($x$,$y$)=($43.25$, $48.5$), 
as expected for a shock. In the post-shock, downstream region, on the upper right side 
of the temperature gradient, the motion of the shocked gas is 
either negligible with the scales plotted ($43.75$, $48.5$) or 
directed outwards of the cluster centre if it is on the subcluster core 
or in its wake. Note that in the rest frame of the shock, 
both  upstream and downstream gas would have inward velocities.

Fig.~\ref{fig:Fig11} is the $z=0.18$ version of 
 Fig.~\ref{fig:Fig10}, computed over a thin slice 
with the same geometry. Here, while the x-range has 
been kept similar as in Fig.~\ref{fig:Fig10}, 
the y-range has been shifted downwards by 1 Mpc to follow the 
propagation of the shock. The same features are apparent, except 
 the velocity-reversing region at the shock which has  
dilated in size along the direction of the propagation of the shock. 
 The upstream accretion velocity field is weaker than at $z=0.20$, a 
possible consequence of the region being farther from the cluster centre. 
 The lower left cold gas core can be seen as the lower temperature zone at ($43$, $48.5$) 
(remember that the colour scale is not the same as for Fig.~\ref{fig:Fig10}). 
The extended shape of the cold gas core suggests that a phenomenon such as adiabatic expansion 
of gas particles heated at core passage begins to develop. The whole cold feature still moves coherently 
with the surrounding flow, as does its wake, down to the 80 kpc resolution mesh on which we computed the velocity field. 
This resolution is sufficient for our purposes as we describe the motions of extended features in 
the ICM but would be too low to draw conclusions on changes in 
the direction and amplitude of the velocity field at the boundary of this cold feature 
as possible precursor of the cold front.   
Given these precautions, the direction of the flow in the zone between the cold zone 
and the shock then slightly rotates to align with the y-axis at $x=43$
and $y=48.25$, while remaining at significant amplitude. At later times 
when the cold front has developed, we have also checked that no 
 velocity discontinuity is apparent on maps of gas velocity similar to Figs.~\ref{fig:Fig10}
 and Figs.~\ref{fig:Fig11}, at the interface between the cold front and its surrounding ICM.

\subsubsection[]{Entropy variation}
\label{sec:Shocks:MegerBow:VelEntropy:Entropy}

The entropy variation due to a single shock or even a series of shocks is too small to be read directly in 
Fig.~\ref{fig:Fig7}, with Mach numbers $M\lsim3$ as typically found for sub-clumps 
orbiting in the ICM.  Therefore, we follow the approach of 
\citet{Ke03} who trace shocks in an SPH simulation of 
the intergalactic medium, and gather 
the evolution of the entropy of a set of gas particles.  
Not only can this method confirm that the large temperature and pressure gradients discussed in 
Section~\ref{sec:Shocks:MegerBow:TempPress} are associated with 
a shock, but it could also be used in higher resolution simulations to 
separate out consecutive shocks, depending on the initial 
set of Lagrangian particles chosen. 

At $z=0.28$, shortly after the merger, we select all gas particles in 
the lower left zone of the ICM, 
with $(x,y,z) \in ([42\;42.5],[47\;48], [48.5\;49.5])\,$Mpc. 
We find about 1100 such particles; they are then tracked as they 
move in the ICM.  This Lagrangian region trails 
the subcluster core \CGOne$\,$ (see the lower left panel of Fig.~\ref{fig:Fig4}) 
and it is initially unaffected by the shock. The possible merger shock that would develop 
upstream of  the subcluster core before first passage as a  result of  
the compression of the intervening gas is expected to have 
little impact on this trailing set of particles, in contrast to the later bow  
shock. In fact, the set of selected particles is right in the direction of 
propagation where the bow shock seems the strongest 
on the $z=0.18$ map of Fig.~\ref{fig:Fig5}. 

\begin{figure}
  \includegraphics[width=\columnwidth]{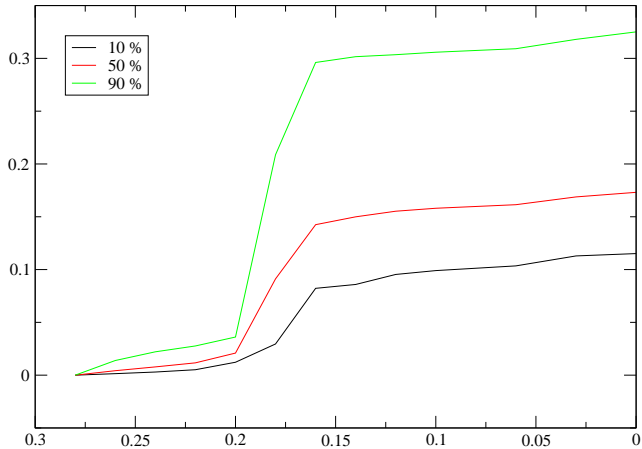}
  \caption{\label{fig:Fig12}The evolution $\Delta\,s_{*}$ of
 the entropy of gas particles selected shortly after the merger. 
The Lagrangian region chosen is a $0.5\,\rmn{Mpc}^{3}$ 
 zone close to the lower left of the 
ICM (trailing \CGOne$\,$ before first core  passage), which is only affected by the bow shock at 
 $z\sim0.18$. We define $s_{*}=\log{(T/\rho^{2/3})}$ and show the 10, 50 
 and 90\% quartiles of the distribution of $\Delta\,s_{*}$ (black, red and green curves 
 respectively).  Note the characteristic synchronous entropy increase between $z=0.2$ 
 and $z=0.18$ due to the shock, together with the spread in $\Delta\,s_{*}$ by more than 
 a factor of 5 between the first and last 10 percents of the distribution.}
\end{figure}

Figure~\ref{fig:Fig12} gives
the entropy evolution $\Delta\,s_{*}$ of this gas, with
$s_{*}=\log{(T/\rho^{2/3})}$.  The black, red and green curves
respectively give the 10, 50 and 90\% quartiles of the
distribution. We distinguish two features. First, there is 
synchronous increase in $s_{*}$ for all particles in $0.2<z<0.16$,
although the degree of variation in $s_{*}$ differs: it is less than
0.05 (Mach number $M_{1}=v_{\tx{shock}}/c_{\tx{s,upstr}}=1.5$) 
for the 10 first percent of the particles and more
than 0.2 ($M_{1}=2$) for the last 10 percent. We have checked that
virtually no selected gas particle is left unaffected. Second, 
when the shock has passed the test Lagrangian region (at $z\lsim0.15$), the 
 entropy per particle changes little down to $z=0$. 
To be concise, and because our goal is to clarify the nature of the temperature gradient 
apparent on the $z=0.18$ maps, we have restricted ourselves 
to only one set of particles of the ICM.

To conclude this paragraph, it is important to note that no gas 
particle of \CGOne$\,$ nor of \CGTwo$\,$ is closely trailing the bow shock 
(see Fig.~\ref{fig:Fig4}), showing that such shocks are only formed because of the 
original internal pressure of the two clumps and because of their supersonic 
velocities (the typical ``piston'' picture) and do not convey  
any elements of the clumps in their vicinity.   

\subsection[]{Secondary merger and bow shocks}
\label{sec:Shocks:SecondaryBow}

Fig.~\ref{fig:Fig5} also shows the presence, after
$z=0.1$, of temperature gradients of smaller size at 
the centre of the cluster. These secondary features, 
probably ``merger'' shocks in the classification of \citet{Mark00}, 
form in a process which is a scaled-down repeat of the initial 
compression of the ICM gas situated between the cold dense subcluster 
cores which took place before their first encounter discussed in
paragraph~\ref{sec:Evol:CentralComp}. At $z=0.1$,
Fig.~\ref{fig:Fig4} shows that both \CDMOne$\,$ and \CDMTwo$\,$ have an extended
shape compared to their $z=0.28$ geometry. The densest parts of the clumps,
however, have almost reached their second encounter in the x-y projection. An
 interesting side effect besides the temperature rise in the
centre of the $z=0.1$ panel of Fig.~\ref{fig:Fig5} is
the thin tail of cold gas leaking from both cold fronts.  It is
particularly visible above the lower left front, and forms in the wake
of the densest part of \CGTwo$\,$ as it falls back towards the cluster
centre.

At $z=0.1$, the secondary  shocks are confined to the very
central region by the convergent gas flows surrounding \CGOne$\,$ and \CGTwo$\,$, as
they move towards their second close encounter, so that their extension is
impeded in the direction of the cold front. They can, however, extend
in the NW-SE direction of the map. As the dense cores go past their
second close encounter this coherent flow stops and the secondary shocks are
released. At this point, they may be more accurately described as ``bow'' shocks 
as the piston effect of \CGOne$\,$ and \CGTwo$\,$ takes place again. The secondary 
shocks clearly extend on at $z=0.06$ just before collapsing
again to the centre at $z=0.03$. 

Although \citet{Naga} state that their bow shocks heat and disrupt their cold front,
it is not clear that this happens here: the secondary bow shocks seem to only affect 
the low temperature tail of the cold fronts which is pointing to the cluster centre, but 
for instance the upper right cold front itself seems largely unaffected
by these secondary features. Mixing or diffusion of the cool phase 
into the hot ICM may explain the final disruption of the cold fronts. 



\section[]{Characterising the cold fronts}
\label{sec:ColdFront}
 
Similarly to what we have previously done for shocks, 
we first demonstrate that the features seen in the 
temperature maps at $z\lsim0.16$ are cold fronts, we then select the particles of the 
lower left cold front and trace them back to elucidate their possible origin.

\subsection[]{Physical attributes}
\label{sec:ColdFront:Attributes}

\begin{figure}
  \includegraphics[width=\columnwidth]{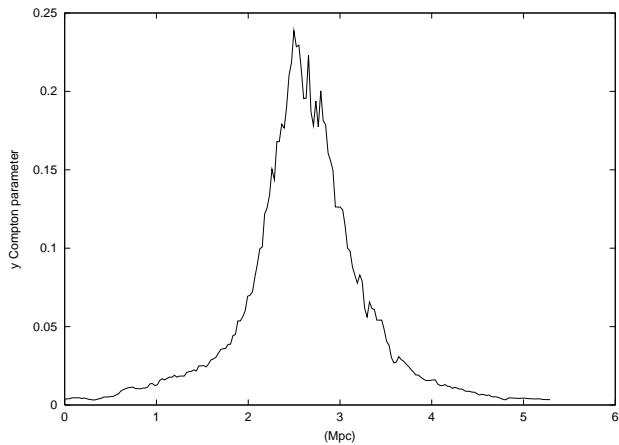}
  \caption{\label{fig:Fig13} Profile of the Compton $y$ parameter of the 
  cluster over a test line passing through the cold fronts, in arbitrary units. The profile has been computed from the value 
 projected over a 500~kpc thick slab normal to the $z$-direction at $z=0.10$, and the  
test line cuts through the centre of the cluster. The label of the abscissa 
has an origin chosen so that the cluster centre is at $\sim2.5$ Mpc. Note that the pressure rise
 to the centre of the cluster is smooth on both sides, and the late-time signature of 
the merging/bow shocks has reached beyond 
the ranges shown here. The complex features seen at $2.5$ Mpc
may be due to the combination of the remains of the cold, high pressure merging sub-clumps 
and the secondary bow shocks.}
\end{figure}

\begin{figure}
  \includegraphics[width=\columnwidth]{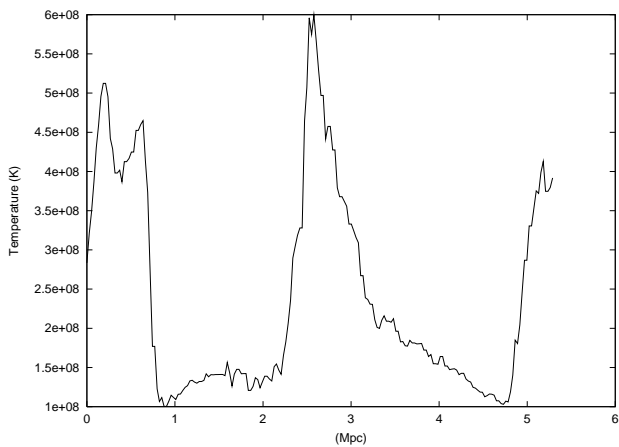}
  \caption{\label{fig:Fig14} 
Same cut as in Fig.~\ref{fig:Fig13}, but for the temperature profile of the gas. 
Note the globally symmetric shape: the gas temperature rises to the left, central and right regions 
($x_{\rmn{loc}}<1$ Mpc, $2<x_{\rmn{loc}}<3$ Mpc, and $x_{\rmn{loc}}>5$ Mpc). 
Between these zones, the signature of the cold fronts and of the wake 
trailing the subcluster cores is striking, as they extend over 
more than 1 Mpc along the direction shown. 
The upper right cold front of the ICM is on the right.}
\end{figure}

Figures~\ref{fig:Fig13} and~\ref{fig:Fig14} 
 respectively give the Compton $y$ parameter, in arbitrary units, 
and temperature profiles computed at $z=0.1$ on a test line 
cutting through approximatively the longest extension of both cold fronts 
and through the centre of the cluster.  
The profiles have been computed from the values projected over a 500~kpc thick slab 
normal to the line of sight (z-direction). The label of the abscissa has an arbitrary origin. 
The pressure profile, on one hand,  rises to the centre of the cluster, is smooth on both sides, and there is 
no discontinuity over the expected location of the cold fronts. The  
complex features at the centre can be due to the presence of high pressure particles 
which have been stripped off the sub-clump cores, of the secondary compression of the 
intervening gas, or of the birth of secondary bow shocks. 
On the other hand, the temperature profile is symmetric but discontinuous. 
The temperature rises in the centre and on the outskirts of the test line, 
but drops by a factor 4 to 6 in $1<x_{\rmn{loc}}<2$ and $3.5<x_{\rmn{loc}}<5$ Mpc  
with respect to the outer temperature value. The exterior discontinuity corresponds to the envelope of the cold fronts.  
At $z=0.1$, \CGOne$\,$ and \CGTwo$\,$ are  $\sim1\,$Mpc apart.  While the  
interior temperature discontinuity at $x_{\rmn{loc}}=2\,$Mpc is sharp and goes over \CGTwo$\,$, 
\CGOne$\,$ is slightly out of the slice resulting in a 
smoother interior temperature discontinuity at $x_{\rmn{loc}}=3.5$ Mpc. Assuming 
that pressure equilibrium has been achieved between the interior of the 
cold front and the surrounding ICM, the factor 3 to 4 drop in local temperature yields 
a factor 3 to 4 increase in the density of gas of the cold front compared to that of the ICM.

Both cold fronts form behind the bow shocks induced by the 
 motion of the cores. While the ICM gas downstream the 
strong shocks cools down as it expands adiabatically, we show in the next paragraph 
that it cannot be responsible for the formation of cold fronts: the 
particles constituting the fronts have a clear-cut origin.

\subsection[]{Tracing the origin of the cold fronts}
\label{sec:ColdFront:Tracing}

We use again the Lagrangian description of the fluid to trace back
the particles of the cold fronts.  We set up a method to select SPH
particles in the simulation directly from both spatial criteria and
from a threshold applied to the fields shown in the maps of 
Fig.~\ref{fig:Fig5} to Fig.~\ref{fig:Fig7}. 
We can combine selection masks along different directions in the case
the target is conveniently expressed as an intersection.

\begin{figure}
  \includegraphics[width=\columnwidth]{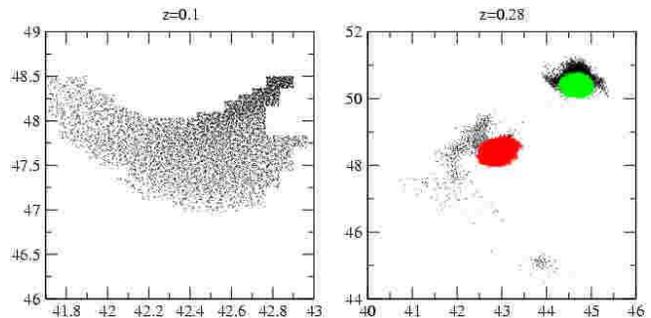}
  \caption{\label{fig:Fig15} Tracing of the gas particles of the lower left cold front. 
 Left panel: position in the x-y plane of the 
  particles of the cold front. The particles have been chosen below an appropriate threshold in the 
  temperature maps at $z=0.10$. Right panel: same  
  particles once they have been traced back to $z=0.28$.  The particles 
of the two cores \CGOne$\,$ and \CGTwo$\,$ are also shown in red and green respectively. 
Note the overlap between \CGTwo$\,$ and the initial position of most of the particles of the cold front. 
A better, high resolution version of this figure is available at: {\tt http://www-astro.physics.ox.ac.uk/$\sim$hxm/ColdFronts/}.}
\end{figure}

To select the cold fronts, we combine two cuts in two
projected, emission-weighted temperature maps along two orthogonal
lines-of-sights through the cluster. Particles of the cold fronts are
selected at $z=0.1$ when the cold fronts are well developed. 
We then trace them back to shortly after $z_{\tx{merg}}$ and
find that almost all of them originate from the two subcluster cores:
gas particles from the lower left cold front come from \CGTwo$\,$ while those
from the upper right cold front come from \CGOne$\,$. Fig~\ref{fig:Fig15} shows the 
origin of the particles selected in this way in the region of the 
lower left cold front, with the particles 
inside the mask shown on the left panel: their position at $z=0.28$ is 
given in black in the right panel, and \CGOne$\,$ and \CGTwo$\,$ are 
overplotted in red and green, respectively. Note the clear overlap between the particles traced back from the 
mask and those of \CGTwo$\,$, supporting that the bulk 
of the mass of the lower left cold front originates in \CGTwo$\,$. There is small additional contribution 
to the mass of the cold front from gas particles which originally surround \CGOne$\,$ at its upper left, 
and from a few ``fuzzy'' particles of the $z=0.28$ ICM.

Conversely, if we follow the forward evolution of \CGOne$\,$ and \CGTwo$\,$ from \zmerg, 
 we find that the envelope of the position of the outer particles of both clumps at $z=0.1$ corresponds  
well with the location of the cold fronts, while their inner densest part has already fallen back 
in the central region of the cluster. This is seen by comparing the 
location of the red clump of the lower left panel of Fig.~\ref{fig:Fig4} 
with the left panel of Fig.~\ref{fig:Fig15}. Apart from the cold front zone, 
neither \CGOne$\,$nor \CGTwo$\,$loose particles to the surrounding ICM. 
Because the evolution is qualitatively very similar for the upper and lower cold front, 
we will exclusively deal with the upper cold front (associated to \CGOne) in the rest of this study. 
 
\begin{figure}
  \includegraphics[width=\columnwidth]{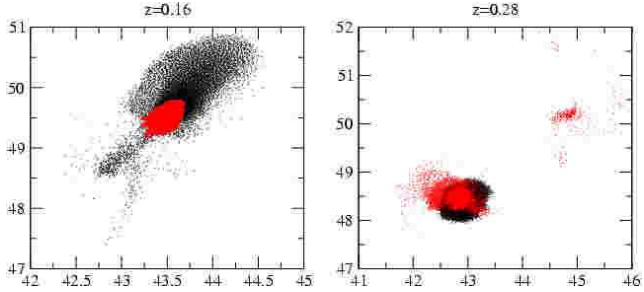}
  \caption{\label{fig:Fig16} Tracing of the high 
pressure gas particles of the upper right cold front region. 
Left panel: projection in the x-y plane of the $z=0.16$ positions 
  of the particles selected at $z=0.1$ by their high Compton $y$ parameter (in red) 
in a $3\times4$ Mpc region encompassing \CGOne: these particles map the bottom 
of the local gravitational potential well.  The $z=0.16$ positions of the particles of $GC1$   
 are repeated in black: note on one side of the potential well 
the extended plume up to $y\sim51$ and $x\sim44.5$ Mpc which constitutes 
the exterior envelope of the cold front,  on the other side a set of particles falling back or trailing. 
Right panel: corresponding positions of the two sets of particles at $z=0.28$. While a fair fraction of the red 
particle overlaps with \CGOne$\,$at $z=0.28$, some surrounding particles, and some particles of the 
lower left of \CGTwo$\,$ also end up with high pressure in the \CDMOne$\,$ potential well at $z=0.1$. 
A better, high resolution version of this figure is available at: {\tt http://www-astro.physics.ox.ac.uk/$\sim$hxm/ColdFronts/}.}
\end{figure}

We show in the next section that the particles of \CGOne$\,$constituting the cold front are expelled from the dark matter potential well,  
while other surrounding and trailing gas particles simultaneously fall in the local potential well. 
 Information about the origin of these latter gas particles 
 is obtained  if we select the particles at $z=0.1$ in the upper cold front region with the highest Compton $y$ parameter. 
The left panel of Fig.~\ref{fig:Fig16} compares the spatial distribution at the epoch 
of formation of the cold front ($z\sim 0.16$) of particles selected according to their 
 $y$ parameter only, in red,  with the positions of particles of \CGOne$\,$ (in black). The right panel traces back both ensembles to 
shortly after \zmerg. While the particles with highest pressure at $z=0.1$ trace as 
expected the minimum of the potential well, they do not map back exactly onto 
 \CGOne: a number of particles also stem from its surroundings and from the outskirt region of \CGTwo$\,$. 
We will see in the next section how an energetic analysis of the black and red particle ensembles of 
Fig.~\ref{fig:Fig16} gives insight into the process at work. 

To summarise, the particles of the cold fronts are essentially composed of
the particles of the cores of the two sub-clumps. Cold front particles 
evolve as a bulk from the epoch of formation of the front until the end of the simulation. 
Interestingly, the gas particles located at the bottom of the 
local gravitational potential associated with the 
\CDMOne$\,$ / \CDMTwo$\,$ substructures 
do not only originate in the cores of the merging clusters but also 
in their initial outskirts at \zmerg.



\section[]{A physical explanation for the formation of cold fronts during mergers}
\label{sec:PhysPic}

To synthesise the description of the previous sections and 
understand in detail the mechanism responsible for the formation of the simulated 
cold fronts, we employ an energetic approach. The variables 
$U$, $W$, $K$ and $T$ respectively stand for the internal, potential, 
kinetic and total energy of a particle. $U$ and $K$ are directly obtained 
from the simulation data, while $W$ is obtained from the contribution of the 
DM particles of the cluster only. We neglect the 
contribution to $W$ from the gas particles, from the large scale structure, 
and from the cosmological constant.  We 
first discuss the energy evolution of the \CGOne$\,$ as 
a whole from \zmerg$\:$down to the present. 
We then focus on the epoch of  formation of the cold front. 
We finally compare to other work.

\subsection[]{Global energetic evolution of \CGOne}
\label{sec:PhysPic:GlobalEnerg}

\begin{figure}
  \includegraphics[width=\columnwidth]{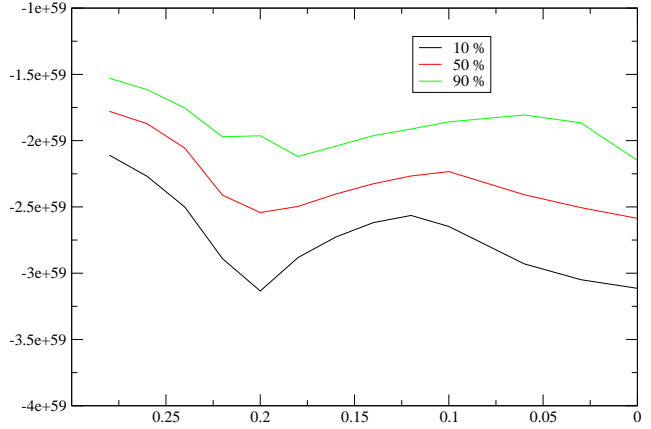}
  \caption{\label{fig:Fig17}Total energy $T$ of the \CGOne$\,$  
particles as a function of redshift, from $z=0.28$. The black, red and green curves respectively give 
 the 10, 50 and 90\% quartiles of the distribution. The abscissa is the y-axis in the simulation  
and the ordinate is the energy (in erg).}	
\end{figure}

Figure~\ref{fig:Fig17} gives the evolution of the distribution of the total energy $T=U+W+K$ of the gas particles 
(of mass $1.3\times 10^{8}\msun$ each) of \CGOne$\,$.  The upper, middle and lower curves 
correspond to the 10, 50 and 90 percent quartiles of the distribution of the particles. 
In practice however, we always have $K<U<\mid W \mid$ for almost all particles, 
and the time evolution of the sum $\sum T$ of the total energies 
of all the particles of \CGOne$\,$ is mainly driven by that of $\sum W$. 

In the first stage from \zmerg $\,$ down to $z=0.2$, the sub-clump core moves towards 
the centre of the newly formed cluster. The total energies $T$ decrease with the gravitational potential energy and 
are not compensated by sufficiently symmetric increase in kinetic energy:   
mostly confined in their local gravitational well, particles of \CGOne$\,$ head towards the cluster centre where the 
 proximity of \CDMTwo$\,$ decreases $W$. In the next stage from $z=0.2$ to $z=0.1$, $T$ globally  
increases as \CGOne$\,$ moves towards its apocentre. Finally, from $z=0.1$ to $z=0$, 
$T$ decreases again as the cores eventually merge. Note that the scatter increases continuously from $z\sim0.15$ 
down to $z=0$: this corresponds to the release of the cold front, which remains stable and does not fall back towards 
the cluster centre with the core of \CGOne$\,$. 

\begin{figure}
  \includegraphics[width=\columnwidth]{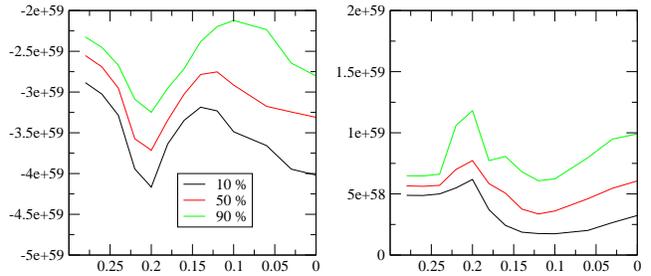}
  \caption{\label{fig:Fig18} Same as Fig.~\ref{fig:Fig17} but for the 
 potential $W$ and internal energy $U$ (left and right panels) of the \CGOne$\,$ 
particles as a function of redshift. 
The black, red and green curves respectively give the 10, 50 and 90\% quartiles of the distribution.}
\end{figure}

Figure~\ref{fig:Fig18} confirms this as it separates the gravitational potential $W$ 
and internal energy $U$ of the particles of \CGOne$\,$ (left and right panels respectively). While 
the evolution of the distribution of the gravitational potential is qualitatively the same as that of $T$, 
the evolution of the internal energy shows that the gas is promptly heated up by compression 
as the gravitational potential decreases around $z=0.2$, part of the gas then cools down 
as a signature of the cold front until $z\sim0.1$. The densest, hottest particles of  \CGOne$\,$ at $z\sim0.1$
 are heated up again during the final inspiralling stage, as shown by the gradient in the 90 percent 
quartile green curve at $z\lsim0.1$. The cold front starts developing at $z\sim0.15$, 
where the width of the distribution of the gravitational energy $W$ of the particles begins to increase down to $z=0$ 
(see the left panel of Fig.~\ref{fig:Fig18}). In fact, gradients in the evolution of the distribution  
of the gravitational energy of the gas particles of \CGOne$\,$ 
are fairly symmetric in the range $z=0.25$ to $z=0.15$  with respect to $z=0.2$, suggesting 
that stripping of the dark matter clump is not dominant. To the opposite, gradients in the evolution of the distribution 
of the internal energy of the gas particles of \CGOne$\,$ are asymmetric: the profiles change in the epochs 
$0.2\gsim z\gsim0.15$ with respect to their pre-encounter evolution. Heating/cooling processes 
are not symmetric with respect to core passage. Particles have been heated up at core passage 
by compression or by possible merging shocks if their density is sufficiently low. 
As the subclump moves on the outward path of its orbit, particles of \CGOne$\,$ 
whose binding energies in the potential well of \CDMOne$\,$ are minimal will (1) 
tend to expand and cool down and (2) more easily decouple from the motion of 
\CDMOne.

Between $z=0.2$ and $z=0.15$, the ``cooling rate'' as given by the gradient of the 10 percent quartile 
of the particle distribution (black line in the right panel of Fig.~\ref{fig:Fig18}) is pronounced. 
Caution is necessary here, given that we simply plot the quartiles of the distribution 
at successive times:  for instance,  
the coolest particles of the distribution at $z=0.15$ can of course be the hottest at $z=0.2$ 
and conversely. After $z=0.1$, the black line (10 percent of the distribution) levels off at 
$\sim2\times 10^{58}$ erg, unlike the red and green lines which show an increase 
in internal energy at $z\lsim 0.12$. The cold front, made of dense particles 
which are expelled out of the local gravitational potential and further cool adiabatically, 
makes a large fraction of the 10 percent particles with internal 
energy below $\sim2\times 10^{58}$ erg. The additional contribution is 
from particles trailing \CDMOne$\,$ and leaking to the centre of the cluster. 



To properly explain the origin of the cold front during mergers requires to understand how a fraction of the \CGOne$\,$ particles 
are able to climb up the local gravitational potential. 

\subsection[]{Formation of the cold front}
\label{sec:PhysPic:FormationColdFront}

\begin{figure}
  \includegraphics[width=\columnwidth]{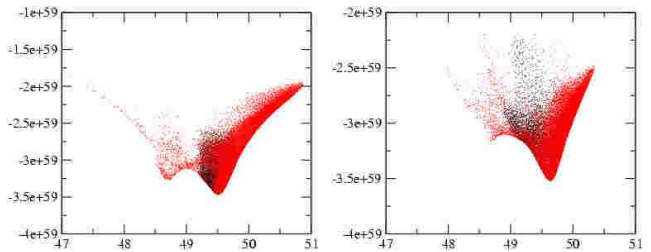}
  \caption{\label{fig:Fig19}Y positions (in Mpc) of the \CGOne$\,$ particles 
  (in red) in the gravitational potential $W$ created by the DM of the whole cluster. The black points give the positions 
  of the pressure-selected particles (by their Compton $y$ parameter at $z=0.1$, see~\ref{sec:ColdFront:Tracing}). 
  The right and left panels correspond to $z=0.16$ and $z=0.12$ respectively.
A better, high resolution version of this figure is available at: {\tt http://www-astro.physics.ox.ac.uk/$\sim$hxm/ColdFronts/}.}
\end{figure}

On Figure~\ref{fig:Fig19}, we plot in red the positions of 
 each particle of \CGOne$\,$ in the gravitational potential $W$, while the black set 
gives the positions of the pressure-selected particles of \CGOne$\,$ 
which will have fallen in the potential well of \CDMOne$\,$ (see~\ref{sec:ColdFront:Tracing}) 
by $z=0.1$. As the red set has been overplotted, 
it can hide black particles. Note also that when a particle belongs to both sets, 
it appears in red.  The positions shown in abscissa correspond to the y-axis 
of the simulation. The right and left panels correspond to $z=0.16$ and $z=0.12$ respectively. 
Recall that \CDMOne$\,$ reaches its apocentre at $z\sim0.15$ (see the 
middle panel of Fig.~\ref{fig:Fig2}). We first describe the 
two panels in turn and then propose an explanation.

At $z=0.16$, the distribution of the core set   
($y>49$ Mpc) of the \CGOne$\,$ particles becomes asymmetric in 
the gravitational potential which is mostly generated by the particles 
of \CDMOne$\,$. We have checked that this asymmetry is not present 
(to this level) at a previous epoch and that it starts  developing at this point. 
At $y\lsim 48.6$, in the wake of the gravitational potential of \CDMOne$\,$ centred at 
$y=49.7$, a second, slightly smaller potential well, which at $z=0.16$ is not yet clearly 
mapped by particles, starts to be filled up with particles of \CGOne$\,$. 
This second potential well is generated by \CDMTwo$\,$ and captures some of the gas particles 
of \CGOne$\,$ which have been stripped off by tidal forces and/or ram pressure. 
These two processes would take place when, 
respectively  \CDMTwo$\,$ is sufficiently close to \CGOne$\,$ and when its 
velocity in the surrounding ICM is high enough. 
Note that a fraction of the black particles have not yet fallen 
in the local potential well of \CDMOne$\,$, even if they stay spatially close to it 
(see the red particles in the left panel of Fig.~\ref{fig:Fig16} for another projection). This 
fraction also corresponds to red particles surrounding \CGOne$\,$ in the right panel 
of Fig.~\ref{fig:Fig16}. 

At $z=0.12$, the cold front is already well developed on the temperature 
maps (Fig.~\ref{fig:Fig5}), and \CDMOne$\,$ has begun its retrograde motion along y. 
The distribution of the \CGOne$\,$ particles in the local 
gravitational potential is now highly asymmetric (see the left panel of Fig.~\ref{fig:Fig19}), 
extending from its minimum at $y=49.5$ to $y\simeq 51$. Particles in the range $50.2<y<51$ 
are responsible for the upper right cold front in the maps. Two other features are striking. First, 
most of the trailing red particles stripped off from \CGOne$\,$ have now fallen back 
in the potential well of the second halo. Second, the  black particles 
have moved close to their selection location as they accreted in the bottom of the \CDMOne$\,$ potential well, where 
they are compressed and sustain the equilibrium with the particles of \CGOne$\,$ forming the asymmetric profile: 
the pressure selection of gas particles in paragraph~\ref{sec:ColdFront:Tracing} has been 
done at slightly later $z=0.1$. In short, we find that the centres of mass of \CGOne$\,$ and \CDMOne$\,$ 
move together after first passage during most of the outward part of their orbit, but 
that a phase shift is introduced as \CDMOne$\,$ reaches its apocentre: the inward motion
 of \CGOne$\,$ is delayed. Similar conclusions have been foreseen by \citet{Roett97} and stressed by  
\citet{Rick01} (see their paragraph 3.2). We now suggest a qualitative picture which accounts for the formation 
of the cold fronts in our simulation.

The effect of the surrounding ICM pressure, present all along the inspiralling 
of the subcluster core, has strong consequences when \CDMOne$\,$ 
reaches the apocentre. In the following qualitative discussion, we will first not 
consider the global pressure gradient of the newly formed cluster, to simplify the picture: we assume 
 a uniform ICM pressure. Note that this hypothesis is a better approximation at the apocentres 
of the orbits of the subclumps, where our cold fronts are generated, 
than at the inner parts of the cluster (see Fig.~\ref{fig:Fig13}, where the 
centre of the cluster is at x$\sim$2.5). We further assume that the flow 
is incompressible, which is valid at the outskirts 
of the orbit where motions are subsonic. Before apocentre 
passage, \CGOne$\,$ stays in phase with \CDMOne$\,$ 
at least from the time of central passage: the offset between 
the two centres of mass is always less than $~100\,$kpc (see Fig.~\ref{fig:Fig2}). 
Recall here that first passage of 
the two subcluster cores occurs at $z\sim0.22$. 
Comparing the red lines (dark matter) in the 
upper left (x-coordinate) and middle (y-coordinate) 
panels of Fig.~\ref{fig:Fig2} between $z=0.22$ and $z=0.16$ 
(the epoch of maximal x and y extension) 
shows that this part of the orbit of \CDMOne$\,$  
is inclined to $\sim$ 60 degrees anticlockwise 
in the x-y plane.

After first core passage, the gravitational force acting on \CDMOne$\,$ 
and \CGOne$\,$ is central, 
and is due to the global smooth cluster potential 
and to the sub-clump 2. However, 
the total force seen by the gas of \CGOne$\,$ also includes 
an ``inertial'' pressure force induced by its accelerated motion,  
which acts in the same direction, but with opposite sign. This 
is true even if the surrounding ICM pressure is uniform. 
(We will employ  acceleration here  in its vectorial sense.)  
Because total forces acting on \CGOne$\,$ and \CDMOne$\,$ differ, 
this will lead to a decoupling of the part of the gas clump (the less bound particles) 
from the dark matter clump. This decoupling of the part of the gas clump 
in turn, is reduced by the gravitational pull introduced by \CDMOne$\,$ as the two components 
of the sub-clump do not anymore share the same spatial distribution. 

Because pressure gradients induced by the accelerated motion of \CGOne$\,$ 
act opposite to the overall gravitational pull, they tend to eject gas 
outside of the local gravitational potential. This liberation of gas is 
in the outward direction with respect to the centre of the cluster. 
It has little effect before apocentre passage because gas particles pushed forward 
in the direction of motion by this force may  immediately fall back in 
the local gravitational potential of the dark matter subclump, whose 
eccentric orbit make it pass over the position of these liberated particles. 
 The potential well of \CDMOne$\,$ will recapture gas particles 
expelled in the direction of its outbound path in a 
process similar to the capture of surrounding gas particles of the ICM 
 in the later inbound motion of \CDMOne$\,$ (see Fig.~\ref{fig:Fig19}). 
This process will not take place at apocentre though, as the 
dark matter subclump \CDMOne$\,$ reaches its extremal x and y-position and 
does not later affect gas particles ejected at this point.

The simulation shows that the restoring gravitational pull that one then expects from the dark matter 
subclump \CDMOne$\,$ is not sufficient to recollect the gas particles 
which have been pushed away at apocentre. 
This is because gas particles can be expelled far enough from the local potential well 
in comparison to its depth (there is more than 0.5$\,$ Mpc
between the bottom  of the local gravitational potential in the right panel of Fig.~\ref{fig:Fig19}). 
In addition, there is no time for a restoring pressure force to develop, as tracing the 
high pressure particles selected at $z=0.12$ (the black set of points in both panels of Figure~\ref{fig:Fig19}) 
shows that a fraction of the surrounding gas particles of \CDMOne$\,$, which are not part of \CGOne$\,$, 
rapidly fall back into the local potential well of \CDMOne$\,$. 
This happens as \CDMOne$\,$ moves back towards the cluster centre and  pressure equilibrium is again achieved 
for the gas trapped in the local potential well of \CGOne$\,$. The particles of \CGOne$\,$ which have been 
left over in an environment of weaker gravitational potential will quickly expand adiabatically to equilibrate with the 
surrounding pressure of the ICM which is much less dense. In this rapid process ($z=0.16\rightarrow 0.12$), 
they cool down from the already low temperature they have as former sub-clump core, 
and appear as cold fronts in the above temperature maps. 
However, the majority of \CGOne$\,$ particles 
are not expelled from the local potential of \CDMOne$\,$: 
 as these particles reach the cluster centre in the second core passage 
they produce the secondary compression/bow shocks.

 We have neglected pressure gradients in this discussion. 
 Pressure gradients will add to the inertial 
effects which we have focused on and will obviously amplify the phenomenon. 
Also, ram pressure stripping will contribute to unbinding some of the gas of \CGOne$\,$ 
when its velocity is large in the surrounding ICM and may also introduce internal motions 
in the dense regions of the gas which can result in patterns observationally very 
similar to cold fronts \citep{He03}. Because our cold fronts do not develop when 
the clump velocity is maximal, and because we do not have sufficient  resolution 
to resolve internal motions in the very core of \CGOne$\,$, it seems unlikely 
that ram pressure effects play a major role are responsible in the formation 
of the cold fronts in our simulations. 

A more detailed, quantitative approach and a precise assessment of the 
relative importance of pressure gradients and ram pressure 
stripping compared to the the ``inertial'' pressure effect 
we discussed will be addressed in future work.

\subsection[]{Comparison to other work}
\label{sec:PhysPic:Compare}

In this paragraph, we briefly put this mechanism 
in the perspective of similar work addressing the formation 
of cold fronts using hydrodynamical 
simulations that has been carried out by \citet{Bia02,Naga,He03}. 

\citet{Bia02} have reached conclusions similar  to ours as they considered the development 
of a cold front in a major cluster merger, arguing  
that adiabatic expansion of the subcluster gas core as it is freed 
from its local gravitational potential could lead to a cold front. 
We have performed a detailed check of their statement that 
this expansion is coupled with compression and heating of the less dense 
cluster gas in forward of the ablated core. Upstream compression and heating are the result 
of the bow shock induced by the motion of the core. Expansion and further cooling 
of part of the dense cold subcluster core gas at the apocentre of its orbit 
lead to the cold front.

\citet{Naga} have shown using temperature and pressure cuts through the ICM 
that the motion of the subcluster core gas in the ambient cluster gas drives a shock 
in front of their simulated cold front during a major merger. 
Confirming that this shock enhances the density 
and temperature gradients between the surrounding ICM 
and the cold gas of the subcluster core, they viewed their cold front as  
that part of this gas which is ``sloshed out'' of its local dark matter 
gravitational potential well. We have clarified this part of the picture 
by tracing back the gas particles of one of 
our simulated cold fronts. According to \citet{Naga}, rapid motion of the cold gas 
inside the freshly shocked gas could result in the sharp boundary. 
Our simulations rather show that 
the role of sloshing in the formation of the cold fronts is significantly 
amplified at the apocentres of the orbits of each of 
the subcluster cores, where our cold fronts develop. 
We have also noted that the velocity of the gas flow over 
the boundaries of our cold front is smooth, confirming 
the claim of \citet{Naga} that the gas 
of the front appears to be moving 
together with its surrounding hot gas. 
This is different from the contact discontinuity model 
discussed by \citet{Vik02} and  
suggests either that there is not enough resolution in 
the simulations or that the \citet{Vik02} picture of cold gas 
surrounded by a uniform wind of hot gas is not fully appropriate.

\citet{He03} have employed 2D eulerian simulations to study  
the evolution of a cold gas 
core as it experiences 
ram pressure stripping when immersed in a background 
uniform flow of hot gas (the flow is switched 
on at the beginning of the simulations).  
Their cold gas core is assumed in hydrostatic 
equilibrium in a local, fixed,  
gravitational potential. \citet{He03} pointed out that gas motions inside the 
cold core are mainly induced by two processes: 
(1) developing Kelvin-Helmoltz 
instabilities, as part of the external regions of the 
cold gas core is driven downstream, 
away from the core as well as (2) the displacement of the 
bulk of the gas core downstream from the centre of the 
local gravitational potential because of ram pressure. Both 
processes result in a vortex forming inside the gas core. 
This vortex brings  low entropy material from the densest parts of the gas core to 
its upstream boundary, where it can adiabatically 
expand and cool, leading to the formation 
of the cold front. \citet{He03} intentionally employed a 
uniform background flow, a fixed local gravitational potential, 
and no  global gravitational potential to simplify the picture 
and focus on the hydrodynamics. In fact, ram pressure 
stripping will also play a role in the development 
of the cold fronts in our simulations. However, we have 
presented evidence that the interplay between the dynamics of the 
subcluster dark matter and gas cores (as embedded in the 
global potential well of the newly formed cluster) might be 
the essential ingredient to the formation of cold fronts in 
massive cluster mergers.


\section[]{Conclusion}
\label{sec:CCL}

Recent high-resolution X-ray observations by \emph{Chandra} 
 have shown that cold fronts are a very common feature of the ICM of bright clusters. 
Cold fronts have been mainly interpreted as a by-product of mergers, 
yet without any clear-cut description of their physical origin. 

We have simulated a late-forming massive merger resulting in a $1.4\times10^{15}\msun$ cluster 
in a concordance cosmological background using a hydrodynamical  
description that is only adiabatic, but 
which is suitable to bring out the Lagrangian origin of the bulk of the cold 
fronts. This sheds light on the details of the formation process of the cold fronts.  

We have first shown that high-resolution SPH simulations can provide 
 temperature maps of an accuracy comparable to that achieved with 
state-of-the-art Eulerian codes, and can be used to study the formation and 
evolution of the ICM during a massive merger. In particular, 
we have singled out the initial compression of the gas intervening 
between the cores of the merging sub-clumps as they move 
towards their first encounter, the generation of two centrally symmetric 
bow shocks as these cores have moved beyond their first passage, 
the outwards propagation of the shock together with the simultaneous formation 
of cold fronts, and the secondary compression of intervening gas before the 
final merging of the cores. Some of the X-ray temperature maps of the 
post-merger ICM are strikingly similar to those 
presented in, e.g. \citet{Mark02b}.   
 
We have then focused on the bow shocks  
sweeping through the ICM, noting that they 
may have interfered with previous shocks developing 
in the central region of the cluster before first core passage. 
We have found that in a 500$\,$kpc thick 
slice roughly containing the plane of merging of the two cores, they affect 
most of the ICM and keep on extending from their 
central generation at $z=0.22$ down 
to $z=0$, first as shocks, later on as pressure waves. 
Eventually they reach beyond the $3.4$~Mpc virial 
radius of the cluster. Even if they weaken from $z=0.12$ on, 
these features can affect the external underdense IGM or the filaments 
and interfere with accretion shocks. They may significantly contribute 
to gas stripping late-type galaxies populating the ICM, and can lead a 
fraction of the ICM to escape the global gravitational potential  
as a result of the pressure wave dissipating energy 
in the outskirts of the cluster. 

Finally, we have considered in detail the formation of 
one of the two symmetric cold fronts. The front develops in the wake of the 
merger/bow shock as some of the cold gas associated with one of 
the dense cores of the sub-clumps which has been heated at 
core passage is freed from its local gravitational potential well.  
This occurs as the sub-clump core reaches the apocentre of its eccentric orbit 
and starts to move back towards second passage. 

At this point, ram pressure stripping is minimal, but the accelerated motion transforms global pressure 
into an effective force pointing outwards, acting on the gas component of the clump in opposite directions to the 
gravitational pull from the clump of dark matter and from the whole cluster. As a result, 
some of this gas is liberated from the local gravitational well, decouples from the dark matter, and quickly cools 
as it expands adiabatically to achieve pressure equilibrium with its surrounding. 

This process has to be compared to, for instance, the pure ram-pressure 
stripping effect advocated by \citet{He03} for the formation of some of the cold fronts. 
Even if ram pressure stripping seems marginal to the formation our two simulated cold fronts, we would like to 
precisely assess the relative importance of the two processes for cold front formation as a function of   
the mass ratio between the merging subclusters in simulations including gas and dark matter in future work.    

Confirming \citet{Naga}, we find that the velocity field is continuous all over the cold front up to the preceding shock, 
 down to the resolution where we have sampled the velocity field ($\sim80$ kpc). 
Higher resolution than we have used is needed to address the question of persistence of  
 the cold front interface against fluid instabilities.  While 
the bow shock/pressure wave keeps on extending, it begins to dissolve at $z=0.12$, however 
our cold fronts remain stable over a longer period, down to $z=0.03$ when 
they start to mix with the hotter ICM. We find that they are little affected by 
the secondary shocks generated at core passage and final inspiralling. 

Finally, SZ polarisation of the CMB combined with the kinetic SZ effect has been brought in  
as a promising tools to constrain the projected velocity of the ICM. We have checked 
in mock SZ polarisation maps that our cold fronts developing on a plane orthogonal to the line of sight 
 do not contribute any significant signal, a consequence of their low bulk velocity with 
respect to the surrounding ICM compared to that of subcluster cores. 

\vspace{0.5cm}

All figures are available at high resolution at the URL : \\
{\tt http://www-astro.physics.ox.ac.uk:$\sim$/hxm/ColdFronts/}

\section*{Acknowledgements}
\label{sec:Ack}

We thank Greg Bryan for useful comments. 
JMD is supported by a Marie Curie Fellowship of the European Community programme 
\emph{Improving the Human Research Potential and Socio-Economic knowledge} 
under contract number HPMF-CT-2000-00967. HM acknowledges financial support from PPARC. 

\bibliographystyle{mnras}

\begin{thebibliography}{19}
\expandafter\ifx\csname natexlab\endcsname\relax\def\natexlab#1{#1}\fi

\bibitem[Bialek et~al.(2002)Bialek, Evrard \& Mohr]{Bia02}
Bialek J.~J., Evrard A.~E., Mohr J.~J., 2002, ApJL, 578, 9

\bibitem[Fujita et~al.(2002)Fujita, Sarazin, Nagashima \& Yano]{Fu02}
Fujita Y., Sarazin C.~L., Nagashima M., Yano T., 2002, ApJ, 577, 11

\bibitem[Heinz et~al.(2003)Heinz, Churazov, Forman, Jones \& Briel]{He03}
Heinz S., Churazov E., Forman W., Jones C., Briel U.~G., 2003, preprint,
  astro-ph/0308131

\bibitem[Keshet et~al.(2003)Keshet, Waxman, Loeb, Springel \& Hernquist]{Ke03}
Keshet U., Waxman E., Loeb A., Springel V., Hernquist L., 2003, ApJ, 585, 128

\bibitem[Markevitch et~al.(2002{\natexlab{a}})Markevitch, Gonzalez, David
  et~al.]{Mark02a}
Markevitch M., Gonzalez A.~H., David L., et~al., 2002{\natexlab{a}}, ApJL, 567,
  27

\bibitem[Markevitch et~al.(2000)Markevitch, Ponman, Nulsen, Bautz, Burke \&
  David]{Mark00}
Markevitch M., Ponman T.~J., Nulsen P.~E.~J., Bautz M.~W., Burke D.~J., David
  L.~P., 2000, ApJ, 541, 542

\bibitem[Markevitch \& Vikhlinin(2001)]{Mark01a}
Markevitch M., Vikhlinin A., 2001, ApJ, 563, 95

\bibitem[Markevitch et~al.(2002{\natexlab{b}})Markevitch, Vikhlinin \&
  Forman]{Mark02b}
Markevitch M., Vikhlinin A., Forman W.~R., 2002{\natexlab{b}}, in { Matter and
  energy in clusters of galaxies\/}, edited by S.~Bowyer, C.-Y. Hwang, vol.~X
  of { ASP conference series\/}

\bibitem[Mazzotta et~al.(2001)Mazzotta, Markevitch, Vikhlinin, Forman, David \&
  VanSpeybroeck]{Maz01}
Mazzotta P., Markevitch M., Vikhlinin A., Forman W.~R., David L.~P.,
  VanSpeybroeck L., 2001, ApJ, 555, 205

\bibitem[Nagai \& Kravtsov(2003)]{Naga}
Nagai D., Kravtsov A.~V., 2003, ApJ, 587, 514

\bibitem[Ricker \& Sarazin(2001)]{Rick01}
Ricker P.~M., Sarazin C.~L., 2001, ApJ, 561, 621

\bibitem[Roettiger et~al.(1997)Roettiger, Loken \& Burns]{Roett97}
Roettiger K., Loken C., Burns J.~O., 1997, ApJS, 109, 307

\bibitem[Springel \& Hernquist(2002)]{SprEntropy02}
Springel V., Hernquist L., 2002, MNRAS, 333, 649

\bibitem[Springel et~al.(2001)Springel, Yoshida \& White]{SprGadget2001}
Springel V., Yoshida N., White S. D.~M., 2001, New Astronomy, 6, 79

\bibitem[Sun et~al.(2002)Sun, Murray, Markevitch \& Vikhlinin]{Sun02}
Sun M., Murray S.~S., Markevitch M., Vikhlinin A., 2002, ApJ, 565, 867

\bibitem[van~de Weygaert \& Bertschinger(1996)]{Wey96}
van~de Weygaert R., Bertschinger E., 1996, MNRAS, 281, 84

\bibitem[Vikhlinin et~al.(2001)Vikhlinin, Markevitch \& Murray]{Vik01}
Vikhlinin A., Markevitch M., Murray S.~S., 2001, ApJ, 551, 160

\bibitem[Vikhlinin \& Markevitch(2002)]{Vik02}
Vikhlinin A.~A., Markevitch M.~L., 2002, AstL, 28, 495

\bibitem[Voit et~al.(2003)Voit, Balogh, Bower, Lacey \& Bryan]{Voi03}
Voit G.~M., Balogh M.~L., Bower R.~G., Lacey C.~G., Bryan G.~L., 2003,
  preprint, astro-ph/0304447

\end{thebibliography}

\bsp
\label{lastpage}
\end{document}